\pdfoutput=1

\documentclass[11pt,twoside,a4paper,cmspaper,final,collab]{cms-tdr}
\def\svnVersion{451105P}\def\cmsCernNoTag{CERN-EP-2017-267}\def\cmsCernDate{\today}\def\cmsMessage{Published in Physics Letters B as \href{http://dx.doi.org/10.1016/j.physletb.2018.02.033}{\doi{10.1016/j.physletb.2018.02.033.}}}

\begin{document}\cmsNoteHeader{BPH-15-005}

\hyphenation{had-ron-i-za-tion}
\hyphenation{cal-or-i-me-ter}
\hyphenation{de-vices}

\RCS$Revision: 438686 $
\RCS$HeadURL: svn+ssh://svn.cern.ch/reps/tdr2/papers/BPH-15-005/trunk/BPH-15-005.tex $
\RCS$Id: BPH-15-005.tex 438686 2017-12-12 21:25:15Z asanchez $

\newcommand{\ltth}{\ensuremath{\lambda_{\theta}^{\mathrm{HX}}}}

\ifthenelse{\boolean{cms@external}}
{\providecommand{\suppMaterial}{the supplemental material
  [URL will be inserted by publisher]}}
{\providecommand{\suppMaterial}{Appendix~\ref{app:suppMat}}\xspace}
\providecommand{\appRef}[2]{\ifthenelse{\boolean{cms@external}}{#1}{#2}}

\cmsNoteHeader{BPH-15-005}
\title{Measurement of quarkonium production cross sections in pp collisions at $\sqrt{s}=13\TeV$}

\date{\today}

\abstract{
Differential production cross sections of prompt $\JPsi$ and
$\psi\mathrm{(2S)}$ charmonium and $\Upsilon\mathrm{(nS)}$ ($\mathrm{n} = 1, 2, 3$) bottomonium
states are measured in proton-proton collisions at
$\sqrt{s} = 13\TeV$, with data collected by the
CMS detector at the LHC, corresponding to an integrated luminosity of 2.3\fbinv for the
$\JPsi$ and 2.7\fbinv  for the other mesons.
The five quarkonium states are reconstructed in the dimuon decay channel, for dimuon
rapidity $\abs{y} < 1.2$. The double-differential cross sections for each state are measured as
a function of $y$ and transverse momentum, and compared to theoretical expectations.
In addition, ratios are presented of cross sections for prompt $\psi\mathrm{(2S)}$ to
$\JPsi$, $\Upsilon\mathrm{(2S)}$ to $\Upsilon\mathrm{(1S)}$,
and $\Upsilon\mathrm{(3S)}$ to $\Upsilon\mathrm{(1S)}$ production.
}

\hypersetup{%
pdfauthor={CMS Collaboration},%
pdftitle={Measurement of quarkonium production cross sections in pp collisions at sqrt (s)  = 13 TeV},%
pdfsubject={CMS},%
pdfkeywords={CMS, quarkonium, cross sections}}

\maketitle

\section{Introduction}

Since the discovery of heavy-quark bound states,
quarkonium production in hadronic collisions has been the subject of many theoretical and experimental studies.
A well established theoretical framework to describe
quarkonium production is nonrelativistic quantum chromodynamics (NRQCD)~\cite{Bodwin,NRQCD1,NRQCD2},
an effective theory that assumes that the mechanism can be factorized in two steps.
In the first step, a heavy quark-antiquark pair is produced in a given spin and orbital angular momentum
state, either in a color-singlet or color-octet configuration.
The corresponding parton-level cross sections, usually called short-distance coefficients (SDCs),
are functions of the kinematics of the state and can be calculated perturbatively, presently up to
next-to-leading order (NLO)~\cite{gong,Kang,bib:BK,bib:Chao}. In the second step, the
quark-antiquark pairs bind into the final quarkonium states through a nonperturbative
hadronization process, with transition probabilities determined by process-independent long-distance
matrix elements (LDMEs). Unlike the SDCs, the LDMEs are presently not calculable and must be obtained
through fits to experimental data~\cite{gong,Kang,bib:BK,bib:Chao,bib:arXiv:1403.3970,Bodwin:2014gia}.
Until recently, for directly produced S-wave quarkonia, the color-octet $^3\mathrm{S}_1$ term was thought
to dominate, which would result in a strong transverse polarization of the mesons relative to their
direction of motion (helicity frame) at large transverse momentum, \pt.

Experiments at the CERN LHC have provided measurements of the production of the S-wave quarkonium states
\Pcgh, \JPsi, \Pgy, and \PgU(nS) ($\mathrm{n} = 1, 2, 3$), and of the P-wave states,
{\ensuremath{{{\chi}_{\mathrm{c1,2}}}}} and
{\ensuremath{{{\chi}_{\mathrm{b1,2}}}}}(1P)~\cite{lhcbchic,lhcbchib,atlaschic,Khachatryan2015383,Chatrchyan2012},
 at center-of-mass energies of 2.76, 7 and 8\TeV. These measurements of the S-wave states include both the
differential cross sections~\cite{lhcbetac,bib:Aad:2011sp,bib:Aad:2014fpa,atlasxs,atlas2016,bib:LHCb1S,bib:LHCb2S,lhcb,bib:ALICE,cmsxs1,cmsxs2,cmsxs7j,cmsxs7u,cmshin1,cmshin2} and polarizations~\cite{alicepol,cmsjpol,cmsypol,lhcbpol,lhcbpol2}, and offer strong indication that, contrary to previous expectations, these mesons are produced unpolarized. Further theoretical
and experimental work can provide deeper insights on how to interpret these observations.
In particular, additional data can help in improving the fits and determine more precisely the relative
weights of the LDMEs.

We report the measurement of double-differential cross sections of five S-wave quarkonium states
 \JPsi, \Pgy, and \PgU(nS) in
\Pp\Pp{} collisions at $\sqrt{s} = 13$\TeV by the CMS detector at the LHC. The increased center-of-mass
energy and production cross sections provide an extended reach in \pt\ and improved statistical
precision relative to similar measurements at 7\TeV~\cite{xsecjpsi2011,cmsxs1,cmsxs2,cmsxs7j,cmsxs7u}.
The measurements performed at 13\TeV also provide the opportunity to test the $\sqrt{s}$ dependence of the
cross sections and to check the validity of the factorization hypothesis and LDME universality implied in NRQCD.

The product of the branching fraction of quarkonia to muon pairs,
$\mathcal{B}(\mathcal{Q} \to \MM)$, and the double-differential production cross section,
$\rd^2\sigma/(\rd\pt\,\rd{}y)$, in bins of \pt and rapidity, $y$, is given by
\begin{equation}
\mathcal{B}(\mathcal{Q} \to \MM) \, \frac{\rd^2\sigma}{\rd\pt\,\rd{}y}= \frac{N(\pt,y)}
{\lumi  \Delta y \Delta \pt} \, \left\langle\frac{1}{\epsilon(\pt,y) \mathcal{A}(\pt,y)}\right\rangle,
\label{eq:xsec}
\end{equation}
where $N(\pt,y)$ is the number of prompt signal events in the bin, \lumi\ is the integrated
luminosity, $\Delta y$ and  $\Delta \pt$ are the bin widths, and
$\left\langle1/(\epsilon(\pt,y) \mathcal{A}(\pt,y))\right\rangle$ represents the
average of the product of the inverse acceptance and efficiency for all the events in the bin.
Only prompt signal events are considered. The nonprompt components of the \JPsi\ and \Pgy\ mesons,
\ie originating from decays of b hadrons, are separated using the decay length defined as
$\ell = L_{xy} \cdot m/\pt$, where $L_{xy}$ is the distance measured in the transverse plane between the
average location of the luminous region and the fitted position of the dimuon vertex,
$m$ is the mass of the \JPsi (\Pgy) from Ref.~\cite{PDG2016}, and \pt the transverse momentum of the dimuon candidate.
For the prompt signal events, we do not distinguish between feed-down decays of heavier quarkonium states and directly produced
quarkonia.

\section{The CMS detector, data set, and event selection}

The analysis uses dimuon events collected in \Pp\Pp{} collisions at $\sqrt{s}=13\TeV$ with the CMS
detector. The central feature of the CMS apparatus is a superconducting solenoid of 6\unit{m}
internal diameter, providing a magnetic field of 3.8\unit{T}. Within the solenoid volume are a
silicon pixel and strip tracker, a lead tungstate crystal electromagnetic calorimeter, and a brass
and scintillator hadron calorimeter, each composed of a barrel and two endcap sections. Forward
calorimeters extend the pseudorapidity ($\eta$) coverage provided by the barrel and endcap detectors. Muons
are detected in gas-ionization chambers embedded in the steel flux-return yoke outside the
solenoid~\cite{cmsmuon}. A more detailed description of the CMS detector, together with a definition
of the coordinate system used and the relevant kinematic variables, can be found in
Ref.~\cite{cmsdet}.

The data were collected using a multilevel trigger system~\cite{cmstrigger}. The first level (L1),
made of custom hardware processors providing coarse momentum information, requires two muons within
the range $\abs{\eta}<1.6$ without requesting an explicit \pt\ threshold on the individual muons.
Second (L2) and third (L3) levels, collectively known as the HLT (High-Level Trigger),
are implemented in software. At these levels, the muon selection is refined, then opposite-charge muon candidates
are paired and required to have an invariant mass in the regions 2.9--3.3, 3.35--4.05, or 8.5--11\GeV for
the \JPsi, \Pgy, and \PgU(nS), respectively.  The dimuon \pt is required to be
above 9.9\GeV for the \JPsi\ and above 7.9\GeV for the remaining states. For all five states,
the dimuon rapidity is restricted to $\abs{y} < 1.25$. A fit of the positions and momenta of the two
muon candidates to a common vertex is performed, and the fit $\chi^{2}$ probability is required to
be above 0.5\%. The sample collected with these triggers has a total integrated luminosity of
$2.3$\fbinv\ for the \JPsi and $2.7$\fbinv\ for the other mesons. The lower value for the \JPsi is the
consequence of the trigger prescaling that was applied to limit the rate during part of the data taking, when
the instantaneous luminosity increased.

When reconstructing the five states offline, further requirements are applied:
only muons with $\pt^{\mu} >$ 4.5 \GeV in the range $|\eta^{\mu}| <$ 0.3, or
$\pt^{\mu} >$ 4.0 \GeV in the range 0.3 $< |\eta^{\mu}| <$ 1.4 are selected.
The muons have to match the triggered pair and be identified as reconstructed tracks with at least five measurements
in the silicon tracker and at least one in the pixel detector. The track is required
to match at least one muon segment identified by a muon detector plane. Loose
criteria are applied on the longitudinal and transverse impact parameters to reject
cosmic rays and in-flight hadron decays. The dimuon vertex $\chi^2$ probability is required to be
greater than 1\%.
In the CMS magnetic field, the two muons can bend towards or away from each
other; only the second type of event is considered in this analysis since the first type exhibits
high trigger inefficiencies.
It was verified that this requirement does not introduce any bias in the determination of the prompt component for the 
\JPsi and \Pgy\ mesons.
The dimuon rapidity is restricted to $\abs{y} < 1.2$.
Trigger bandwidth limitations prevented the extension of the measurement to the full CMS acceptance.

The double-differential cross sections  are presented in four (two) rapidity bins for the prompt
\JPsi and \Pgy\ (\PgU(nS)), and in several bins of \pt, covering a \pt range between 20 and
120 (100)\GeV  for \JPsi(\Pgy , \PgU(nS)), extending up to 150 (130) \GeV for measurements
integrated in rapidity.

\section{Acceptance and efficiencies}\label{sec:acc}

The acceptance is calculated using simulated events produced with a single-particle event
generator. The quarkonium states are generated with a flat $y$ distribution and a realistic \pt
distribution derived from data~\cite{cmsxs2,cmsxs7j}, covering the analysis phase space.
The \PYTHIA 8.205~\cite{Pythia} Monte Carlo event generator is used to produce an unpolarized
dimuon decay (corresponding to a flat dimuon angular distribution), also accounting for final-state
photon radiation.
The simulated events include multiple proton-proton interactions
in the same or nearby beam crossings (pileup), with the distribution
matching that observed in data, with an average of about 11 collisions per bunch crossing.
The acceptance for events in a given (\pt,$\abs{y}$) range is defined as
the ratio of the number of generated events that pass
the kinematic selection criteria described above to the total number of simulated events in that \pt
and $\abs{y}$ range. The acceptance depends on the quarkonium polarization. It is derived for
the unpolarized scenario, which is compatible with experimental measurements within uncertainties.
We also calculate multiplicative correction factors that allow,
from the unpolarized case, to infer the acceptance that corresponds to three different values of the
polar anisotropy parameter, \ltth,  in the helicity frame:
$-1$ (fully longitudinal),  $+1$ (fully transverse), and $k$, with $k$ reflecting the
CMS measured value of \ltth{} for each quarkonium
state~\cite{cmsjpol,cmsypol}, also used in Refs.~\cite{cmsxs7j,cmsxs7u}.
The multiplicative factors to convert the cross sections
calculated using the unpolarized scenario to the ones calculated employing one of the polarization
scenarios described above are provided.
It was verified that the use of only events with two muons bending away from each other does not introduce any bias in the determination of the acceptance.

The single-muon trigger, reconstruction, and identification efficiencies are measured
individually from data as a function of muon \pt and $\abs{\eta}$, applying a
tag-and-probe~\cite{cmsxs1,xsecjpsi2011} technique on \JPsi and \PgUa\ candidates acquired with triggers that are independent from those used for the  measurements of the yields.
The individual efficiencies are multiplied and then parameterized using a sigmoid function.
The dimuon efficiency is obtained as the product of the efficiencies of the two muons,
multiplied by a correction factor, $\rho$, that takes into account the correlation between
the two muons.
The $\rho$ factor is derived from data, using a trigger, independent from the ones used for
the measurement of the yield, requiring a single muon at L1. $\rho$ becomes increasingly important with higher dimuon \pt,
when the two muons are close to each other in space, causing the efficiency to decrease.
Dimuon efficiencies are around 85\% for the
\JPsi and \Pgy\ up to a dimuon \pt of 50\GeV and decrease slowly for higher \pt due to
the $\rho$ factor. In the case of the \PgU(nS) states, the dimuon efficiencies are nearly
constant around 90\%.
The acceptance and efficiency term in Eq. (\ref{eq:xsec}) is obtained
by averaging the values of the inverse of the acceptance times efficiency for all
the individual dimuon candidates in each \pt and $\abs{y}$ range.

\section{Determination of the yields}

The signal and background yields are obtained through an extended unbinned maximum-likelihood
fit to the dimuon invariant mass distribution in the case of
the \PgU(nS) states, and to the dimuon invariant mass and decay length distributions for the \JPsi and \Pgy\
mesons. In both cases, the number of signal and background candidates are free parameters in
the fit.

The three \PgU(nS) signal peaks are modeled with Crystal Ball (CB) functions~\cite{bib:CrystalBall},
 composed of a Gaussian core, characterized by a mean $m$, a width $\sigma_m$, and a tail characterized by
two parameters, $n$ and $\alpha$. The CB function is used to account for the energy
loss due to the final-state radiation of the muons.
The mean mass values are fixed to those of the Particle Data Group~\cite{PDG2016},
multiplied by a common factor that calibrates the mass scale, left as a free parameter in the fit.
The width of the CB function is a free parameter only in the case of the \PgUa, while the width
of the CB functions describing the \PgUb\ and \PgUc\ peaks are fixed to the width
of the \PgUa, scaled by the ratio of their masses to the mass of the \PgUa. The \PgU(nS) dimuon mass resolution $\sigma_m$ is a
function of rapidity and spans the range 60 to 90~\MeV for $\abs{y} < 1.2$ in the case of the \PgUa.
The tail parameters $n$ and $\alpha$ are the same for all three CB functions; $n$ is fixed and
$\alpha$ is constrained to a Gaussian probability distribution. Both constraints are derived from
a fit of the \PgUa\ dimuon invariant mass shape, using the \pt-integrated distribution to reduce
the statistical fluctuations. The background is modeled using an exponential function.

For the \JPsi\ and \Pgy\  mesons, an additional nonprompt component originating from the decay
of b hadrons must be taken into account. The prompt and nonprompt yields are measured by
fitting the dimuon invariant mass and
decay length distributions. The \JPsi\ dimuon invariant mass distribution is modeled by the sum of
a CB and a Gaussian function with common mean, while the corresponding \Pgy\ distribution is
described using only a CB
function. The widths of the CB and Gaussian functions, as well as the $\alpha$ of the CB functions,
are free parameters. The $\sigma_m$ varies as a function of rapidity between 20
and 50 (40)~\MeV for the \JPsi(\Pgy) state. The $m$ and $n$ parameters are fixed to values
derived from fits to the invariant mass distribution of the \pt-integrated data.
An exponential is used to describe the dimuon mass background.
The decay length distribution
is modeled by a prompt signal component represented by a resolution function, a nonprompt term given by an
exponential function convolved with the resolution function, and a background term represented
by the sum of a resolution function plus an exponential decay function to take into
account prompt and nonprompt background components.
The resolution function is modeled by the sum of two Gaussian functions whose widths are taken as
the event-by-event decay length uncertainty, multiplied by global scale factors.
The two scale factors are free parameters in the fit and are constrained with Gaussian probability
distributions that are derived
from fits to the \pt-integrated data, less affected by statistical fluctuations. The effective width of
the two Gaussian functions is approximately 25\mum.

To verify that the fits to the quarkonium states are unbiased and the uncertainties are
correctly modeled, 1000 pseudo-experiments were produced from simulation.
Similarly, simulated events were used to test the hypotheses made on the constraints of the parameters.
Differences in the event-by-event uncertainty information between signal and background candidates could
introduce biases in the fitting of the decay length using the simplified model described above,
but we verified that these effects are negligible.
Examples of fits to the invariant mass and decay length distributions are provided
in Figs.~\appRef{1--2}{\ref{fig:fitjpsi}--\ref{fig:fitupsn}} of \suppMaterial.

\section{Systematic uncertainties}\label{sec:systematics}

Systematic uncertainties are due to the measurement of the integrated luminosity
(2.3\%)~\cite{lumi_moriond}, the determination of the signal yields, and the dimuon efficiencies
and acceptances. Uncertainties in the estimation of the yields are evaluated by changing the
signal and background models used in the maximum-likelihood fits. To assess the systematic
uncertainty in the modeling of the signal invariant mass distribution of
each state, the $n$ and $\alpha$ parameters of the CB function are varied by up to $\pm 5$ standard deviations,
one at a time, while the mean, which is constrained in the nominal fit, is allowed to float.
The half-differences between the largest resulting deviations of the signal yields measured
in the fit from the nominal yields are added in quadrature to obtain an uncertainty
in the modeling of the signal.
The systematic uncertainty originating from a possibly imperfect description of the background is
evaluated by changing the background model from an exponential to a linear function.
The observed differences from the nominal signal yields are taken as a systematic uncertainty.
The total uncertainty in the determination of the yields is obtained as the sum in quadrature
of the uncertainties in signal and background, and is about 2.0\% for all quarkonium states.

Uncertainties in the discrimination between charmonia that are promptly produced rather than
originating from b hadron decays arise from the determination of the primary vertex position
(the production point of the mesons, which enters in the calculation of the decay length)
and from the modeling of the signal and background in the decay length distributions.
We assess the uncertainty originating from the choice of the primary vertex by
using an alternative to the average position of the luminous region, the position of the
collision vertex closest to the dimuon vertex extrapolated towards the beam line.
The systematic uncertainty related to the description of the background
is evaluated by measuring the difference
between the prompt fractions using the nominal fit and a fit modeling the background
by the sum of four exponential functions and a simplified resolution function composed
of only one single Gaussian function.
To study the impact of imperfect modeling of the resolution function, the scale parameters of
the Gaussian functions that had Gaussian constraints in the nominal fit are varied by $\pm 1$ standard deviation.
Similarly, we assess the impact of modeling the nonprompt signal by fixing the parameterization
of the  exponential decay function. The systematic uncertainty stemming from the choice of the
primary vertex is added in quadrature with the uncertainty derived from the fit strategy. The
latter is calculated as half of the difference between the maximum deviations
observed from the  nominal fit when the above variations are applied one by one. The total systematic
uncertainty in the determination of the nonprompt yield is less than 3\% in almost all the
(\pt, $y$) bins for the \JPsi meson, without a dominant contribution from any one of the
sources described above.
The largest systematic uncertainty in the \Pgy\ measurements can reach a maximum of 16\%,
mostly owing to the uncertainty in the modeling of the background decay length distribution.
The effect of pileup on the analysis results has been studied using both data and
simulation, and found to be negligible.

Uncertainties in the single-muon efficiencies, reflecting their statistical precision as well as
possible imperfections of the parametrization, are evaluated by varying the three parameters
of the sigmoid function used to parameterize the single-muon efficiencies
within their uncertainties.
The resulting systematic uncertainties are nearly constant as a function of \pt and
are around 2.5\% for the \JPsi\ and \Pgy, and 1.8\% for the \PgU(nS) in the central
regions $\abs{y} < 0.6$ and around 1\% in the remaining rapidity regions.
The L3 single-muon efficiencies are calculated from simulations because of the low number of
collected events useful for their measurements. The corresponding uncertainty is estimated to be 3\%.

Systematic uncertainties related to the $\rho$ factor are of three kinds. The first
originates from the number of events available in the control sample collected with the independent trigger
used to evaluate the $\rho$ factor. The relative uncertainty is about 1\% from 20 to 50\GeV
and increases to about 5\% near 100\GeV, with no dependence on rapidity.
The measurement of the $\rho$ factor also requires the evaluation of an additional single-muon
efficiency using  the tag-and-probe method, which introduces an uncertainty of about 1\% at low
\pt (below 50\GeV) and up to 4\% at high \pt. Moreover, we assign the fractional difference in
the $\rho$ factor obtained from data and simulation as a systematic uncertainty. The difference
is in the range 2--5\% up to 60\GeV and increases slowly for higher \pt, reaching a value of up
to 15\%, in the worst case. This is the dominant uncertainty for all the quarkonium states
except the \Pgy .

The finite number of events generated for the acceptance calculation imposes a systematic
uncertainty of 0.5\% at low \pt and up to 6\% at high \pt. Other sources of systematic
uncertainties, like the kinematic modeling of simulated events, are found to have a negligible
influence on the acceptance calculation. The effect of the quarkonium polarization on the acceptance
is not treated as a systematic uncertainty; instead correction factors are provided in
\suppMaterial to recalculate the cross sections according to different
polarization scenarios.

For the cross sections measured in the rapidity-integrated range $\abs{y} < 1.2$, we
conservatively assign the total systematic uncertainties of the most-forward rapidity range,
which are larger than the uncertainties for central rapidities. Taking advantage of the
larger yields in the integrated-rapidity range, an additional \pt bin was added for each state.
The systematic uncertainty in the yields for this bin was evaluated as described above for the
other bins, while for other uncertainties the same value as in the neighboring lower-\pt bin
was used. It was verified that systematic uncertainties extrapolated to the additional \pt bin
have either negligible \pt dependence in that region or are negligibly small compared to other
systematic or statistical uncertainties.

For the measurement of the ratios of the cross sections of the prompt
\Pgy, \PgUb, and \PgUc\ states relative to
their ground states, the systematic uncertainties in the yields, the $\rho$
factor, the single-muon efficiencies, and the acceptance are the only ones considered.
Uncertainties in the yields for the ratio of \Pgy\ and \JPsi cross sections are treated as uncorrelated,
because their corresponding yields are determined from independent fits.
In contrast, yield uncertainties are treated as correlated for the ratio of the \PgU(nS) to \PgUa\ cross sections,
as they are extracted from a combined fit to the three states, as shown in Fig.~\appRef{2}{\ref{fig:fitupsn}} of \suppMaterial.
The correlation factors are found to be approximately 5\%, causing no significant effect on the final systematic uncertainty.
The same single-muon efficiencies are used for all the measured cross sections,
therefore their uncertainties are treated as correlated in all the ratios.
The systematic uncertainties in the ratios are determined by consistently
varying the efficiencies in the numerator and the denominator
by their uncertainties and recalculating the ratios.
The resulting effect is less than 0.4\%.
The uncertainty in the integrated luminosity is fully correlated, and is not included in the ratios.
Uncertainties in the $\rho$ correction factor are treated as uncorrelated.

The statistical uncertainty in the \Pgy\ to \JPsi cross section ratio is more important than any systematic uncertainty
except for the high-\pt region, where the $\rho$ factor uncertainty is the dominant one, reaching 28\%.
For the \PgUb\ to \PgUa\ and  \PgUc\ to \PgUa\  cross section ratios, the uncertainty in the $\rho$ factor dominates
across the entire \pt region, ranging from 3\% to 12\%.

\section{Results}

\begin{figure}[htb]
\centering
\includegraphics[width=0.48\textwidth]{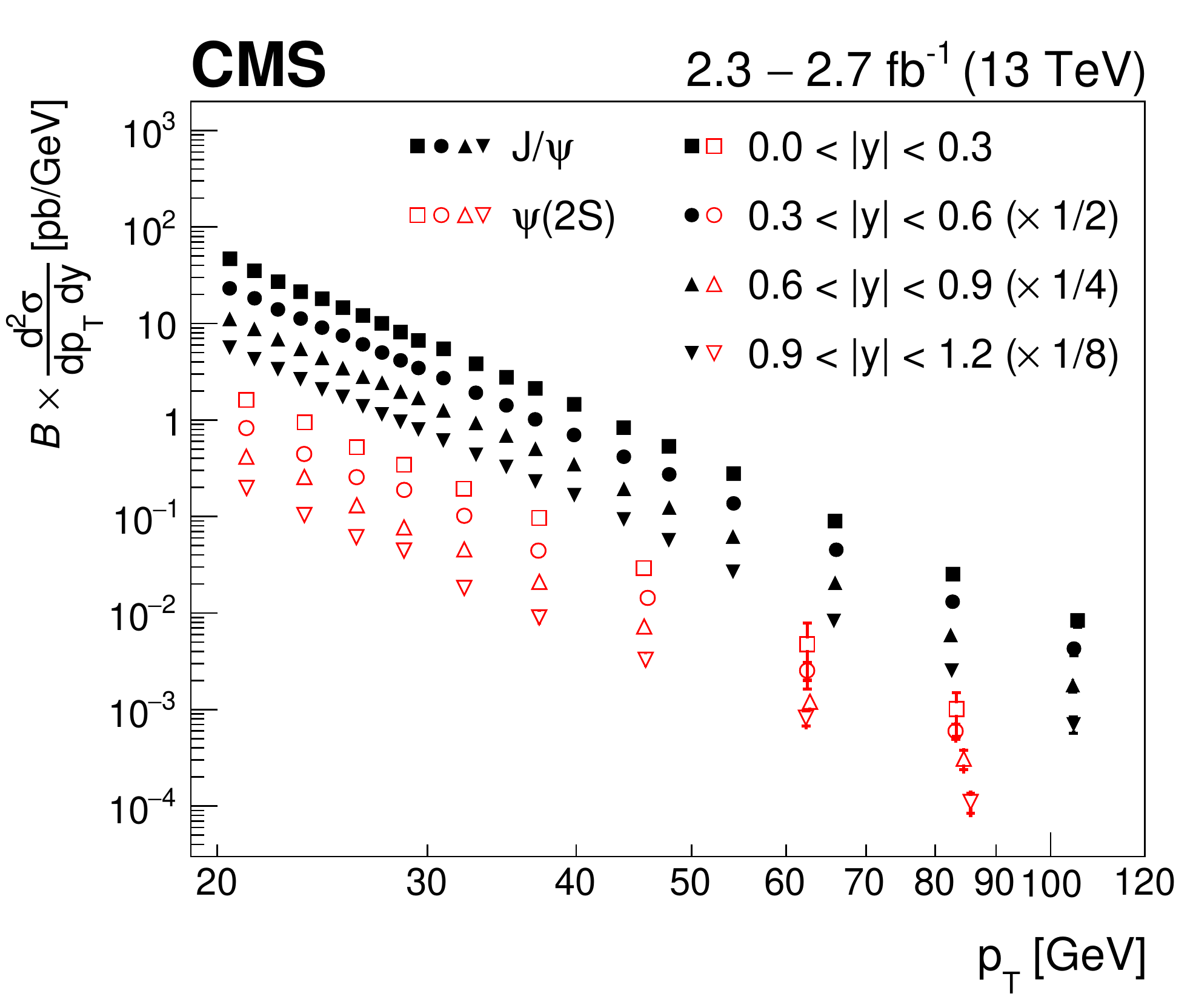}
\includegraphics[width=0.48\textwidth]{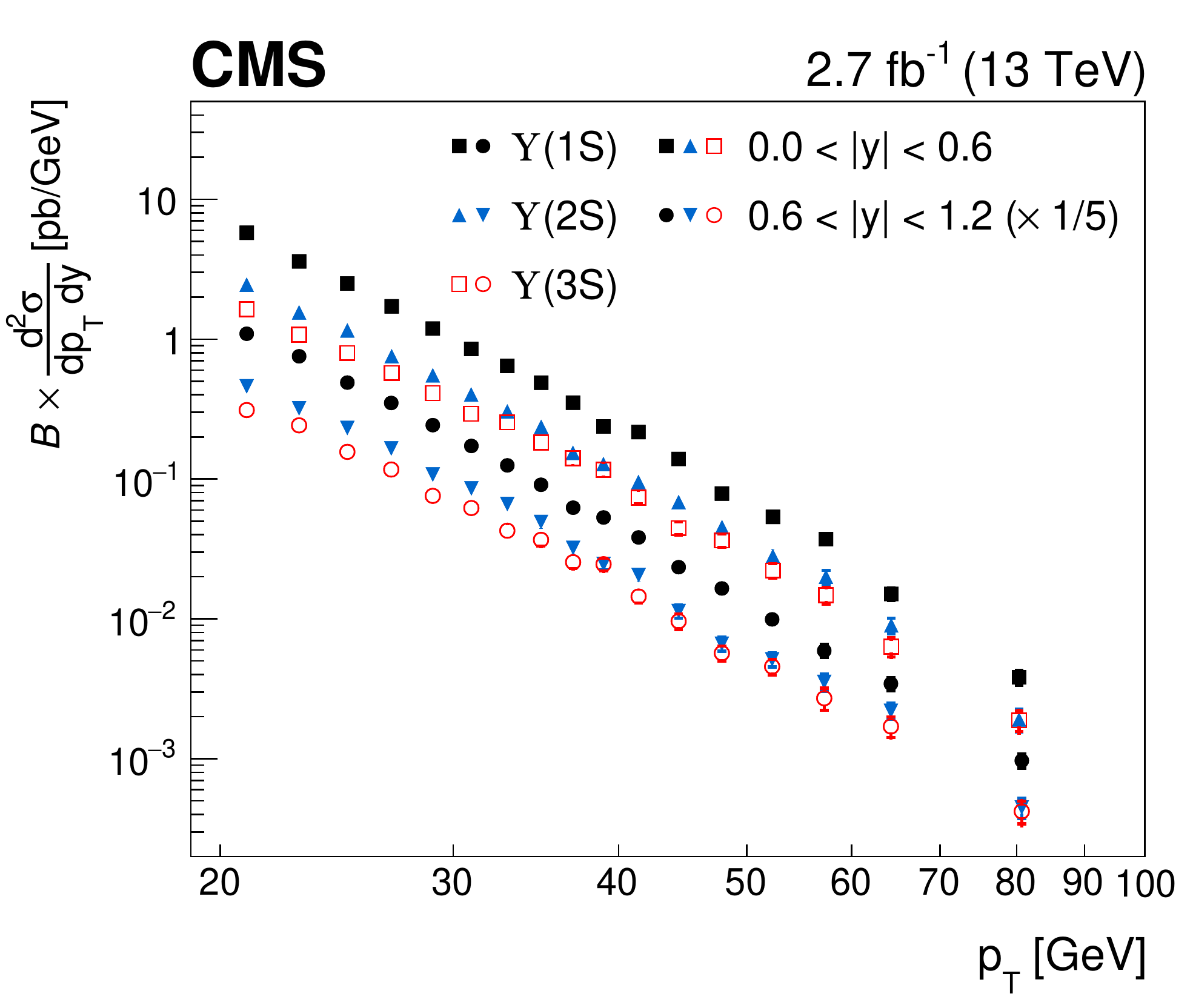}
\caption{
The product of the measured double-differential cross sections and the dimuon branching fractions for prompt
\JPsi\ and \Pgy\ (left) and the \PgU(nS) (right) mesons as a function of \pt, in four and two
rapidity regions, respectively, assuming unpolarized dimuon decays. For presentation
purposes, the individual points in the measurements are scaled by the factors given in
the legends. The inner vertical bars on the data points represent the statistical uncertainty,
while the outer bars show the statistical and systematic uncertainties, not including the
2.3\% uncertainty in the integrated luminosity, added in quadrature. For most of the data points,
the uncertainties are comparable to the size of the symbols. 
The data points are shown at the average \pt in each bin.
}
\label{fig:diffxsec}
\end{figure}

\begin{figure*}[htb]
\centering
\includegraphics[width=0.48\textwidth]{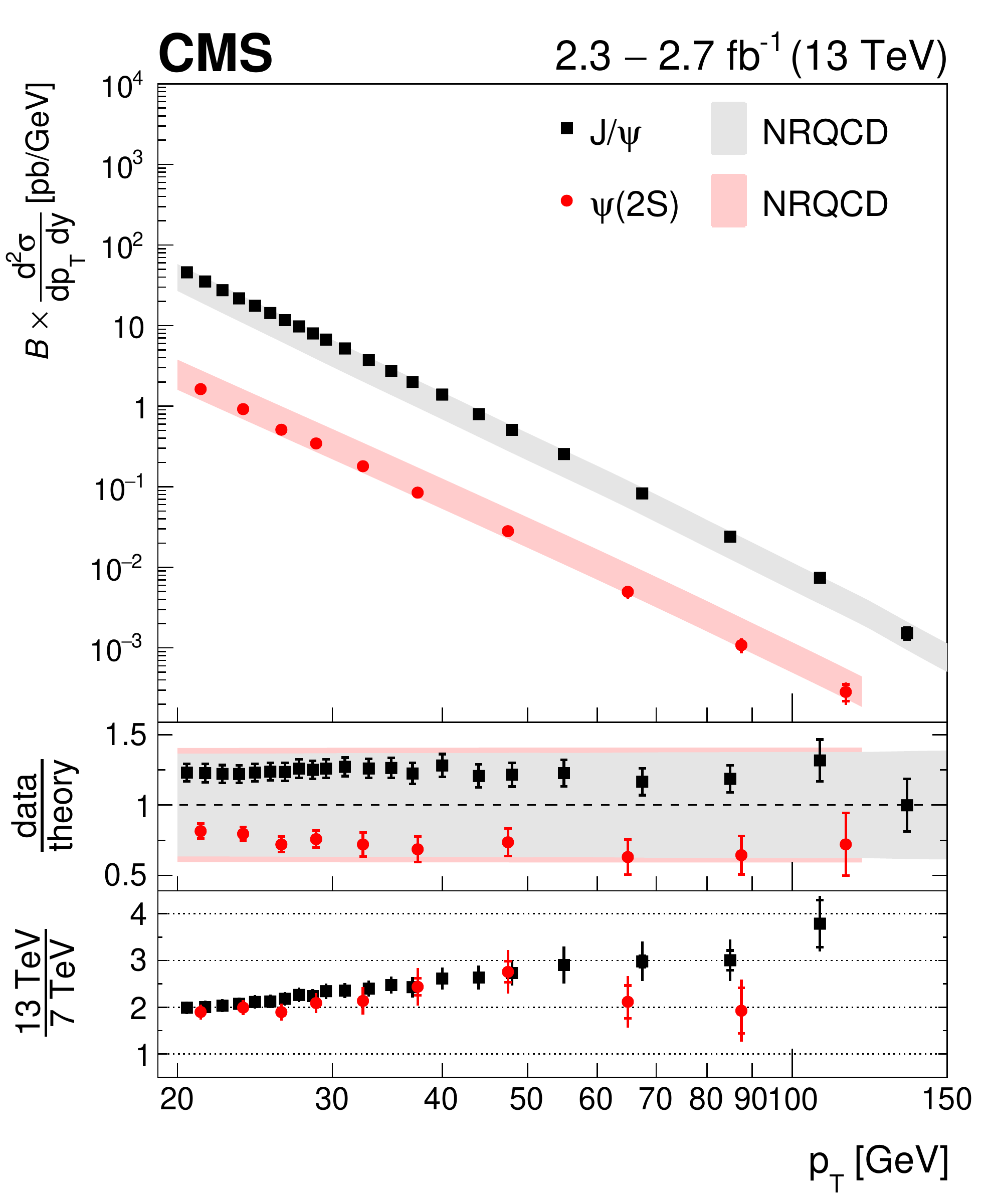}
\includegraphics[width=0.48\textwidth]{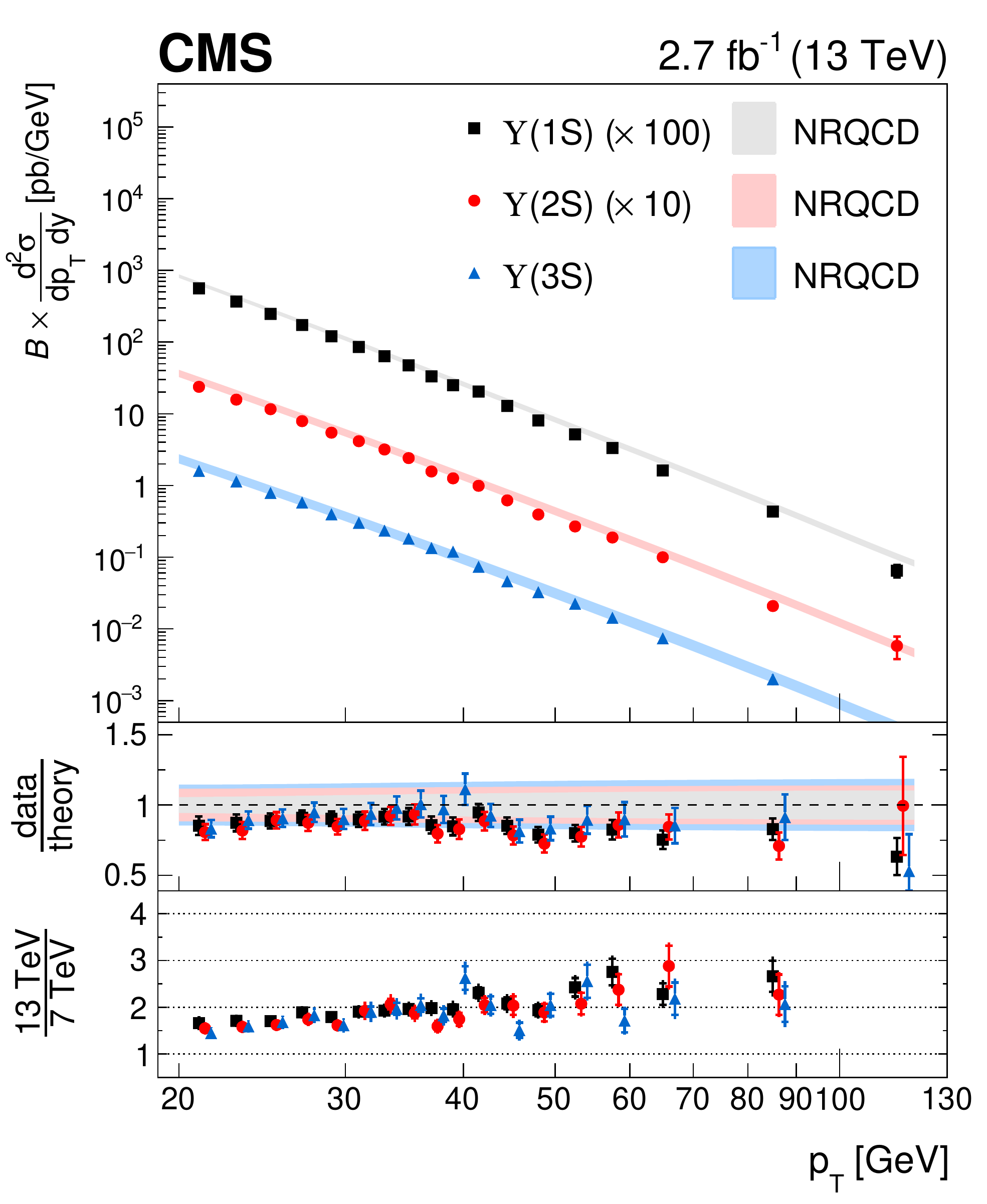}
\caption{
The measured double-differential cross sections times branching fractions of the prompt \JPsi and \Pgy\ (left)
and the \PgU(nS) (right) mesons (markers), assuming unpolarized dimuon decays, as a function
of \pt, for $\abs{y} < 1.2$, compared to NLO NRQCD predictions~\cite{predictions1,predictions2}
(shaded bands). The inner vertical bars on the data points represent the statistical
uncertainty, while the outer bars show the statistical and systematic uncertainties,
including the integrated luminosity uncertainty of 2.3\%, added in quadrature. The middle
panels  show the ratios of measurement to theory, where the vertical bars depict the total
uncertainties in the measurement. The widths of the bands represent the theoretical
uncertainty, added in quadrature with the uncertainties in the dimuon branching
fractions~\cite{PDG2016}. The lower panels show the ratios of cross sections
measured at $\sqrt{s} = 13$\TeV to those measured at 7\TeV~\cite{cmsxs7j,cmsxs7u}.
All uncertainties in the 7 and 13\TeV results are treated as uncorrelated.
The data points are shown at the average \pt in each bin.
}
\label{fig:xsecs}
\end{figure*}

The measured double-differential cross sections times the dimuon branching fractions are presented in
Fig.~\ref{fig:diffxsec} as a function of \pt, for four rapidity ranges in the case of the
prompt \JPsi\ and \Pgy\ states, and two rapidity ranges for the \PgU(nS). The top panels of
Fig.~\ref{fig:xsecs} show the measured cross sections times branching fractions for the
rapidity-integrated range $\abs{y}<1.2$. The presented results are obtained under the assumption
of unpolarized production, which is very close to the polarization that was measured by
CMS~\cite{cmsypol,cmsjpol}.
If the quarkonium states are fully polarized,  the cross sections
can change by up to 25\%. The numerical values of the cross sections for all five quarkonium states in the
chosen bins of \pt and $\abs{y}$ in the unpolarized scenario are reported in
Tables~\appRef{1--5}{\ref{tab:jpsixsec}--\ref{tab:y3sxsec_eta1}} of \suppMaterial.
Tables~\appRef{6--10}{\ref{tab:poljpsi}--\ref{tab:polups3}} list the multiplicative scale factors
needed to recalculate the cross sections
in the three different polarization scenarios described in Section~\ref{sec:acc}. The conversion
to a new polarization scenario is achieved by multiplying the unpolarized cross section result in
each (\pt, $\abs{y}$) bin by the corresponding scale factor.

The NLO NRQCD predictions~\cite{predictions1,predictions2} are in agreement
with the measured cross sections times branching fractions within uncertainties, as shown in
the top panels of Fig.~\ref{fig:xsecs}. The ratios of the measured to predicted values
are plotted in the middle panels of Fig.~\ref{fig:xsecs}, where the vertical bars represent
the experimental uncertainties. The shaded bands show the theoretical uncertainties stemming
from the extraction of the LDMEs, renormalization scales, and the choice of c and b quark
masses, added in quadrature with the uncertainties in the dimuon branching fractions~\cite{PDG2016}.
The theory tends to underestimate (overestimate) the cross section for the \JPsi\ (\Pgy), while
staying within the one-standard-deviation uncertainty band. The bottom panels of
Fig.~\ref{fig:xsecs} show the ratios of the \pt\ differential cross sections times branching
fractions measured at $\sqrt{s} = 13$\TeV and 7\TeV~\cite{cmsxs7j,cmsxs7u} for $\abs{y} < 1.2$.
The 13\TeV cross sections of all five quarkonium states are factors of 1.5 to 3 larger than
the corresponding 7\TeV cross sections, changing slowly as a function of dimuon \pt. An
increase of this order is expected from the evolution of the parton distribution functions.

\begin{figure}[htb]
\centering
\includegraphics[width=0.48\textwidth]{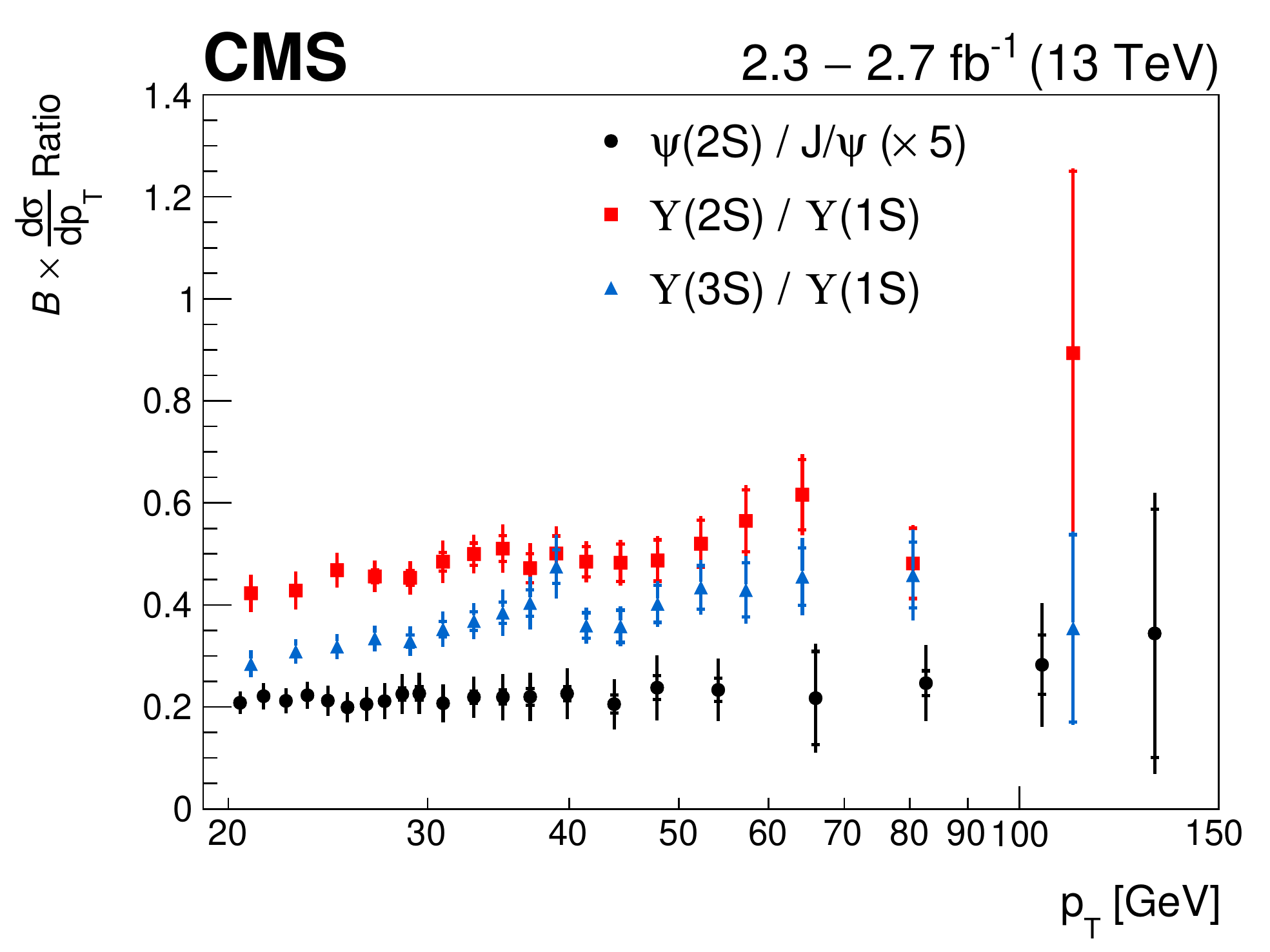}
\caption{
Ratios of the \pt differential cross sections times dimuon branching fractions of the
prompt \Pgy\ to \JPsi, \PgUb\  to \PgUa, and \PgUc\  to \PgUa\ mesons for
$\abs{y} < 1.2$. The inner vertical bars represent the statistical uncertainty, while the
outer bars show the statistical and systematic uncertainties added in quadrature.
The ratio of the \Pgy\ to \JPsi\ meson cross sections is multiplied by a factor 5 for better
visibility.
}
\label{fig:oniarat}
\end{figure}

Figure~\ref{fig:oniarat} shows the production cross sections times dimuon branching fractions
of the radial excitations relative to the ground state in the charmonium and bottomonium
systems for $\abs{y} < 1.2$. The prompt \Pgy\ to \JPsi\ meson cross section ratio is constant as a
function of \pt, while the cross sections of the excited \PgU\ states relative to the \PgUa\
show a slight increase with \pt. The numerical values of these ratios are reported in
Table~\appRef{11}{\ref{tab:ratio_excited}} of \suppMaterial.

\section{Summary}

The double-differential production cross sections of the \JPsi, \Pgy, and \PgU(nS) ($\mathrm{n} = 1, 2, 3$)
quarkonium states have been measured, using their dimuon decay mode, in \Pp\Pp{} collisions
at $\sqrt{s} = 13$\TeV with the CMS detector at the LHC. The production cross sections of all
five S-wave states are presented in a single analysis.
The measurement has been performed as a function of transverse momentum (\pt) in several bins of
rapidity ($y$), covering a \pt range 20--120\GeV for the \JPsi\ meson and 20--100\GeV for the remaining states.
The cross sections integrated over $\abs{y} < 1.2$ are also presented, and extend the \pt
reach to 150 and 130\GeV, respectively.
Also presented are the ratios of cross sections measured at
$\sqrt{s} = 13$ (this analysis) and 7\TeV (from Refs.~\cite{cmsxs7j,cmsxs7u}),
as well as the cross sections of the prompt \Pgy, \PgUb, and \PgUc\ mesons relative to their
ground states. These results will help in testing the underlying hypotheses of
nonrelativistic quantum chromodynamics and in providing further input to constrain the theoretical
parameters.

\clearpage
\begin{acknowledgments}
The authors would like to thank Yan-Qing Ma for providing theoretical calculations of the cross
sections shown in Fig.~\ref{fig:xsecs}.

We congratulate our colleagues in the CERN accelerator departments for the excellent performance of the LHC and thank the technical and administrative staffs at CERN and at other CMS institutes for their contributions to the success of the CMS effort. In addition, we gratefully acknowledge the computing centers and personnel of the Worldwide LHC Computing Grid for delivering so effectively the computing infrastructure essential to our analyses. Finally, we acknowledge the enduring support for the construction and operation of the LHC and the CMS detector provided by the following funding agencies: BMWFW and FWF (Austria); FNRS and FWO (Belgium); CNPq, CAPES, FAPERJ, and FAPESP (Brazil); MES (Bulgaria); CERN; CAS, MoST, and NSFC (China); COLCIENCIAS (Colombia); MSES and CSF (Croatia); RPF (Cyprus); SENESCYT (Ecuador); MoER, ERC IUT, and ERDF (Estonia); Academy of Finland, MEC, and HIP (Finland); CEA and CNRS/IN2P3 (France); BMBF, DFG, and HGF (Germany); GSRT (Greece); OTKA and NIH (Hungary); DAE and DST (India); IPM (Iran); SFI (Ireland); INFN (Italy); MSIP and NRF (Republic of Korea); LAS (Lithuania); MOE and UM (Malaysia); BUAP, CINVESTAV, CONACYT, LNS, SEP, and UASLP-FAI (Mexico); MBIE (New Zealand); PAEC (Pakistan); MSHE and NSC (Poland); FCT (Portugal); JINR (Dubna); MON, RosAtom, RAS, RFBR and RAEP (Russia); MESTD (Serbia); SEIDI, CPAN, PCTI and FEDER (Spain); Swiss Funding Agencies (Switzerland); MST (Taipei); ThEPCenter, IPST, STAR, and NSTDA (Thailand); TUBITAK and TAEK (Turkey); NASU and SFFR (Ukraine); STFC (United Kingdom); DOE and NSF (USA).

\hyphenation{Rachada-pisek} Individuals have received support from the Marie-Curie program and the European Research Council and Horizon 2020 Grant, contract No. 675440 (European Union); the Leventis Foundation; the A. P. Sloan Foundation; the Alexander von Humboldt Foundation; the Belgian Federal Science Policy Office; the Fonds pour la Formation \`a la Recherche dans l'Industrie et dans l'Agriculture (FRIA-Belgium); the Agentschap voor Innovatie door Wetenschap en Technologie (IWT-Belgium); the Ministry of Education, Youth and Sports (MEYS) of the Czech Republic; the Council of Science and Industrial Research, India; the HOMING PLUS program of the Foundation for Polish Science, cofinanced from European Union, Regional Development Fund, the Mobility Plus program of the Ministry of Science and Higher Education, the National Science Center (Poland), contracts Harmonia 2014/14/M/ST2/00428, Opus 2014/13/B/ST2/02543, 2014/15/B/ST2/03998, and 2015/19/B/ST2/02861, Sonata-bis 2012/07/E/ST2/01406; the National Priorities Research Program by Qatar National Research Fund; the Programa Severo Ochoa del Principado de Asturias; the Thalis and Aristeia programs cofinanced by EU-ESF and the Greek NSRF; the Rachadapisek Sompot Fund for Postdoctoral Fellowship, Chulalongkorn University and the Chulalongkorn Academic into Its 2nd Century Project Advancement Project (Thailand); the Welch Foundation, contract C-1845; and the Weston Havens Foundation (USA).
\end{acknowledgments}
\bibliography{auto_generated}
\ifthenelse{\boolean{cms@external}}{}{
\clearpage
\appendix
\numberwithin{table}{section}
\numberwithin{figure}{section}
\section{Dimuon invariant mass and lifetime distributions, numerical values of differential cross section,  and correction factors for alternative polarization scenarios \label{app:suppMat}}
\providecommand{\pbGeV}{\ensuremath{\mathrm{pb}/\GeVns}\xspace}
\newcommand{\ltt}{\ensuremath{\lambda_{\theta}}}
\begin{figure*}[htbp]
\centering
\includegraphics[width=0.40\textwidth]{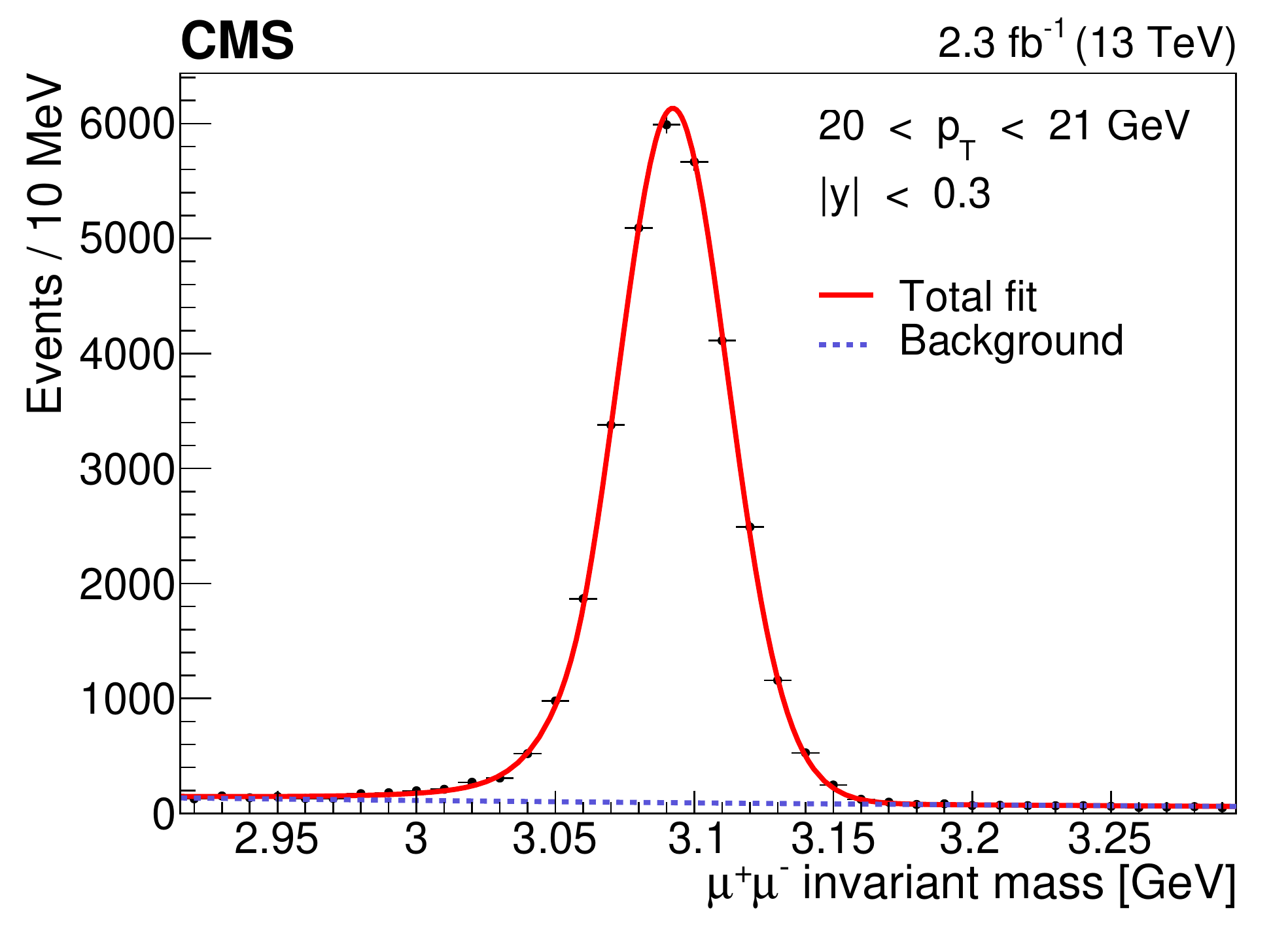}
\includegraphics[width=0.40\textwidth]{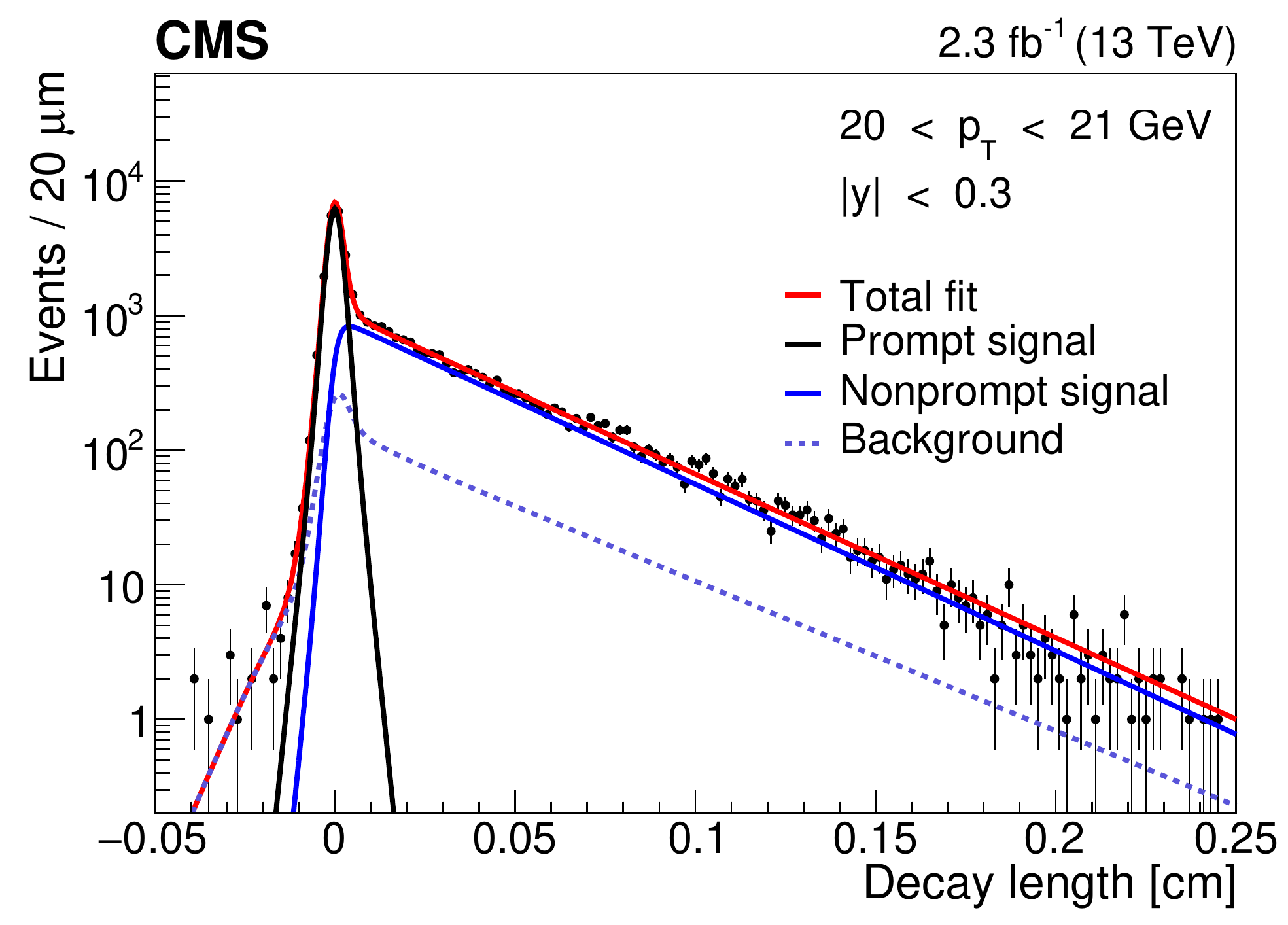}
\includegraphics[width=0.40\textwidth]{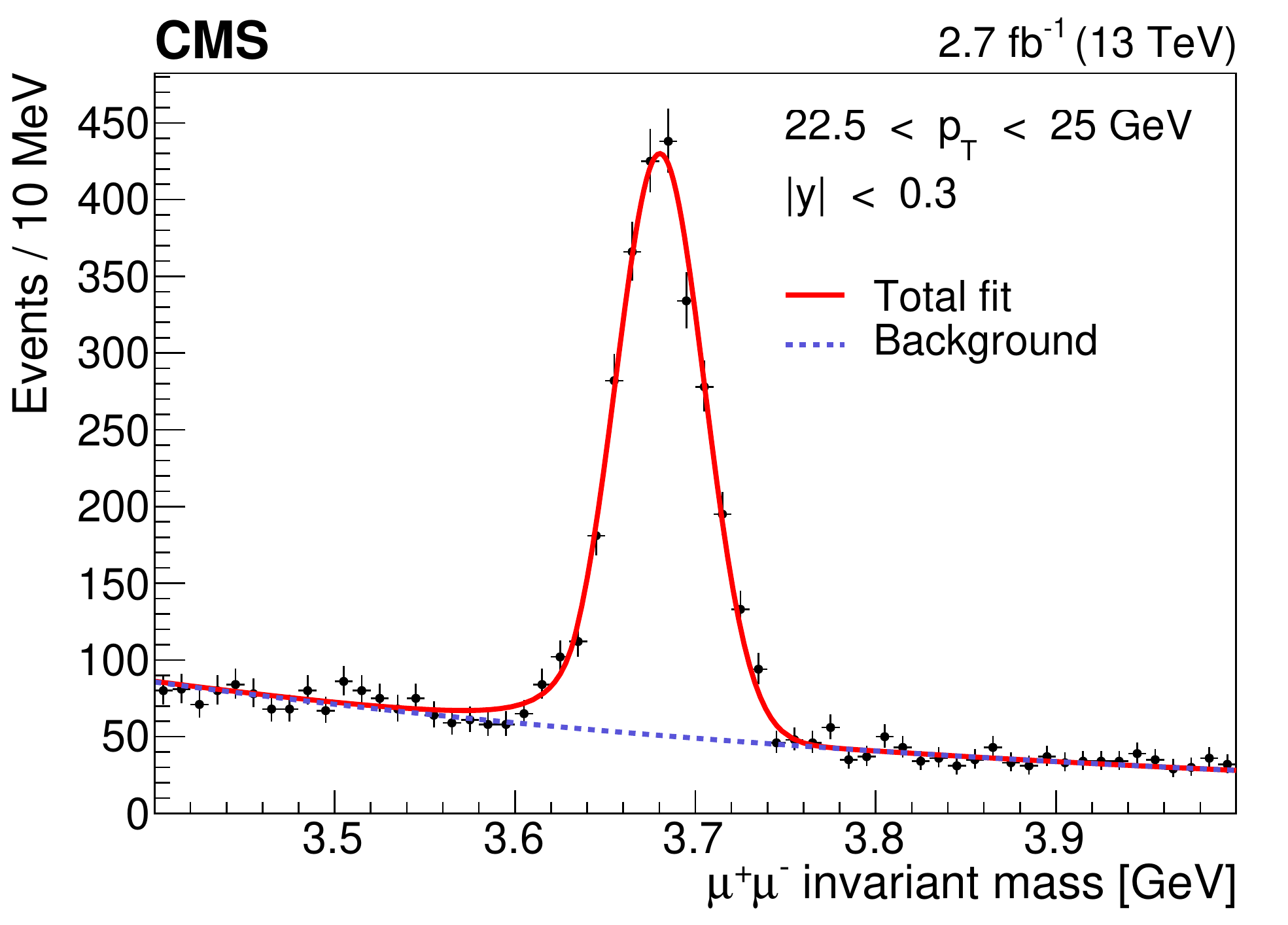}
\includegraphics[width=0.40\textwidth]{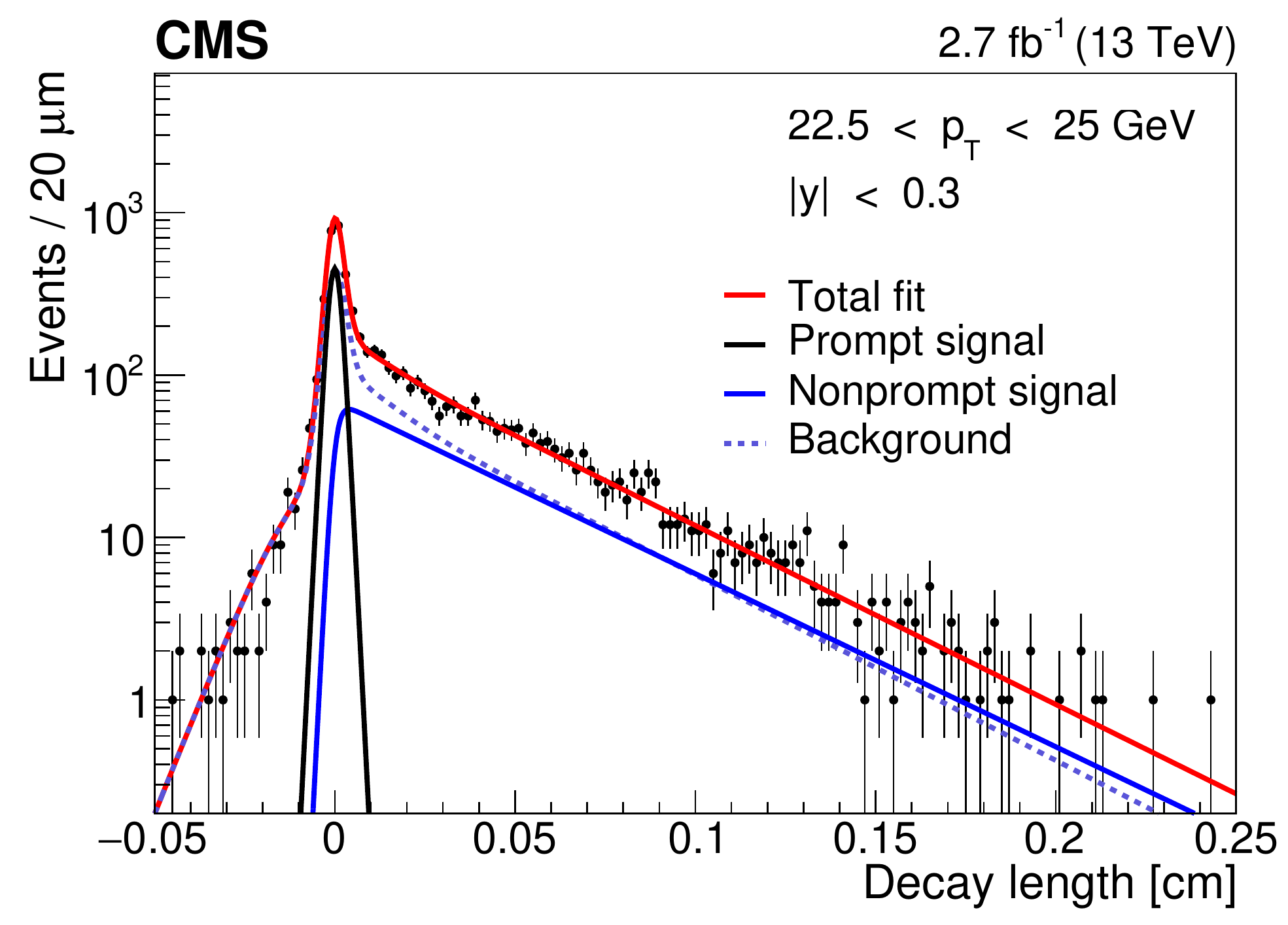}
\caption{Examples of fits of the dimuon invariant mass (left) and decay length (right) distributions
for \JPsi (upper row) and \Pgy\ (lower row) candidate events in the \pt and $\abs{y}$ ranges given in the plots.
The results from the total fit and from the various components included in the fit are shown.
}
\label{fig:fitjpsi}
\end{figure*}

\begin{figure}[htb]
\centering
\includegraphics[width=0.42\textwidth]{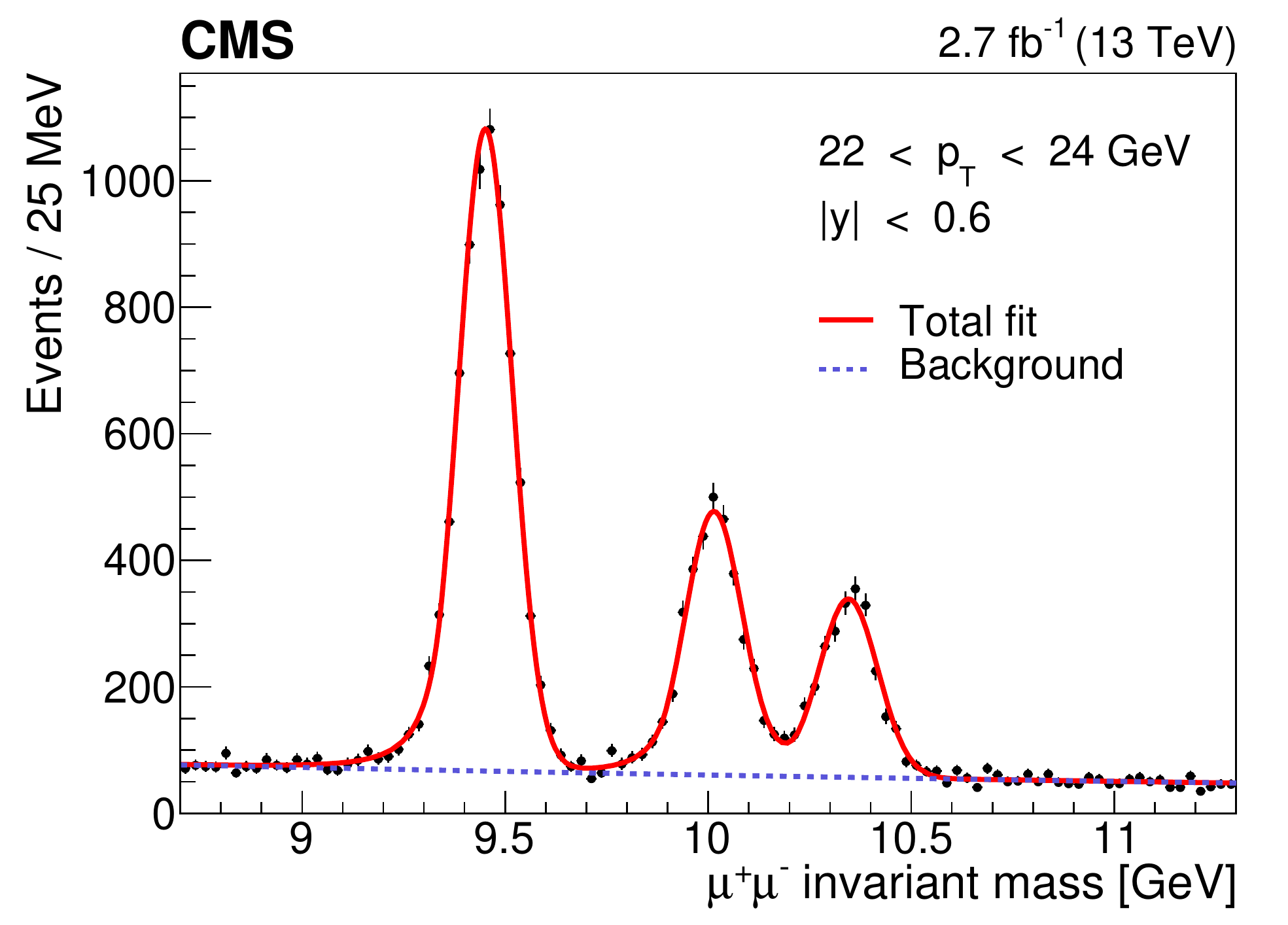}
\caption{Examples of a fit of the dimuon invariant mass distribution for the \PgU(nS) candidate events in the
\pt and $\abs{y}$ ranges given in the plot. The results from the total fit and for the background
component are shown.
}
\label{fig:fitupsn}
\end{figure}

\clearpage

\setlength\tabcolsep{4.0pt}
\begin{sidewaystable}[htb]
\centering
\topcaption{Double-differential cross section times the dimuon branching fraction of the \JPsi meson for different ranges of \pt,
in bins of $\abs{y}$ and for the full $\abs{y}$ range, for the unpolarized decay hypothesis, with their statistical and
systematic uncertainties in percent. The average \pt value in each bin is also given. The global uncertainty in
the integrated luminosity of 2.3\% is not included in the systematic uncertainties.
}
\label{tab:jpsixsec}
{
\small
\begin{tabular}{cc|ccc|ccc|ccc|ccc|ccc}
                            &                         &  \multicolumn{15}{c}{${\cal B} \, \rd\sigma^2/\rd{}\pt \rd{}y$}  \\
\multicolumn{1}{c}{\pt}  & $\langle{\pt}\rangle$ & \multicolumn{3}{c}{$\abs{y} < 0.3$}  & \multicolumn{3}{c}{$0.3 < \abs{y} < 0.6$}  &  \multicolumn{3}{c}{$0.6 < \abs{y} < 0.9$}  & \multicolumn{3}{c}{$0.9 < \abs{y} < 1.2$} & \multicolumn{3}{c}{$\abs{y} < 1.2$} \\
\cline{3-17}
[\GeVns{}]          &  [\GeVns{}]  & [\pbGeV{}]& stat \%  & syst \% & [\pbGeV{}]& stat \%  & syst \% & [\pbGeV{}]& stat \%  & syst \% & [\pbGeV{}]& stat \%  & syst \% & [\pbGeV{}]& stat \%  & syst \% \\
\hline
 20--21 &  20.5 & 4.68E+01 & 1.7 & 5.3 & 4.63E+01 & 1.3 & 4.6 & 4.47E+01 & 1.2 & 4.5 & 4.51E+01 & 1.3 & 4.6 & 4.58E+01 & 0.7 & 4.6 \\
 21--22 &  21.5 & 3.52E+01 & 1.3 & 5.4 & 3.65E+01 & 1.2 & 4.8 & 3.52E+01 & 1.2 & 4.6 & 3.42E+01 & 1.3 & 4.8 & 3.53E+01 & 0.6 & 4.8 \\
 22--23 &  22.5 & 2.72E+01 & 1.4 & 5.2 & 2.80E+01 & 1.3 & 4.5 & 2.75E+01 & 1.3 & 4.4 & 2.69E+01 & 1.3 & 4.6 & 2.74E+01 & 0.7 & 4.6 \\
 23--24 &  23.5 & 2.14E+01 & 1.5 & 5.0 & 2.25E+01 & 1.4 & 4.5 & 2.18E+01 & 1.4 & 4.3 & 2.12E+01 & 1.5 & 4.4 & 2.18E+01 & 0.7 & 4.5 \\
 24--25 &  24.5 & 1.80E+01 & 1.6 & 5.0 & 1.81E+01 & 1.5 & 4.5 & 1.76E+01 & 1.5 & 4.3 & 1.66E+01 & 1.6 & 4.5 & 1.76E+01 & 0.8 & 4.5 \\
 25--26 &  25.5 & 1.46E+01 & 1.8 & 5.0 & 1.50E+01 & 1.7 & 4.5 & 1.38E+01 & 1.7 & 4.3 & 1.39E+01 & 1.8 & 4.5 & 1.43E+01 & 0.9 & 4.5 \\
 26--27 &  26.5 & 1.21E+01 & 1.9 & 5.1 & 1.22E+01 & 1.8 & 4.4 & 1.13E+01 & 1.9 & 4.3 & 1.11E+01 & 1.9 & 4.5 & 1.17E+01 & 0.9 & 4.5 \\
 27--28 &  27.5 & 1.00E+01 & 2.1 & 5.0 & 1.00E+01 & 2.0 & 4.4 & 9.76E+00 & 2.0 & 4.3 & 9.17E+00 & 2.1 & 4.5 & 9.75E+00 & 1.0 & 4.5 \\
 28--29 &  28.5 & 8.14E+00 & 2.3 & 5.1 & 8.31E+00 & 2.2 & 4.5 & 7.88E+00 & 2.2 & 4.3 & 7.67E+00 & 2.3 & 4.5 & 7.99E+00 & 1.1 & 4.5 \\
 29--30 &  29.5 & 6.68E+00 & 2.5 & 5.2 & 6.92E+00 & 2.4 & 4.5 & 6.78E+00 & 2.4 & 4.4 & 6.39E+00 & 2.5 & 4.6 & 6.70E+00 & 1.2 & 4.5 \\
 30--32 &  31.0 & 5.47E+00 & 1.9 & 5.2 & 5.44E+00 & 1.9 & 4.5 & 5.03E+00 & 1.9 & 4.3 & 4.91E+00 & 2.0 & 4.6 & 5.20E+00 & 1.0 & 4.6 \\
 32--34 &  33.0 & 3.84E+00 & 2.3 & 5.4 & 3.84E+00 & 2.2 & 4.6 & 3.72E+00 & 2.2 & 4.4 & 3.50E+00 & 2.3 & 4.8 & 3.72E+00 & 1.1 & 4.7 \\
 34--36 &  35.0 & 2.78E+00 & 2.7 & 5.7 & 2.84E+00 & 2.5 & 4.9 & 2.76E+00 & 2.5 & 4.9 & 2.62E+00 & 2.7 & 5.1 & 2.75E+00 & 1.3 & 5.0 \\
 36--38 &  37.0 & 2.12E+00 & 3.1 & 6.2 & 2.03E+00 & 2.9 & 5.4 & 2.02E+00 & 3.0 & 5.3 & 1.85E+00 & 3.1 & 5.6 & 2.00E+00 & 1.5 & 5.5 \\
 38--42 &  39.8 & 1.45E+00 & 2.6 & 6.4 & 1.40E+00 & 2.5 & 5.7 & 1.39E+00 & 2.5 & 5.6 & 1.33E+00 & 2.6 & 5.9 & 1.39E+00 & 1.3 & 5.8 \\
 42--46 &  43.8 & 8.33E-01 & 3.3 & 6.9 & 8.33E-01 & 3.2 & 6.1 & 7.74E-01 & 3.3 & 5.9 & 7.47E-01 & 3.4 & 6.5 & 7.96E-01 & 1.7 & 6.2 \\
 46--50 &  47.8 & 5.34E-01 & 4.2 & 7.0 & 5.48E-01 & 4.0 & 6.1 & 4.96E-01 & 4.2 & 6.1 & 4.54E-01 & 4.5 & 6.7 & 5.08E-01 & 2.1 & 6.3 \\
 50--60 &  54.2 & 2.79E-01 & 3.7 & 7.8 & 2.74E-01 & 3.5 & 7.1 & 2.49E-01 & 3.7 & 7.1 & 2.14E-01 & 4.1 & 7.7 & 2.54E-01 & 1.9 & 7.2 \\
 60--75 &  66.0 & 8.96E-02 & 5.4 & 8.0 & 9.05E-02 & 5.0 & 6.9 & 8.23E-02 & 5.6 & 6.9 & 6.64E-02 & 6.2 & 8.2 & 8.28E-02 & 2.7 & 7.3 \\
 75--95 &  82.7 & 2.54E-02 & 9.0 & 7.7 & 2.62E-02 & 8.5 & 6.4 & 2.37E-02 & 8.7 & 6.4 & 2.03E-02 & 9.6 & 7.6 & 2.39E-02 & 4.4 & 6.3 \\
 95--120 &  104.7 & 8.37E-03 & 15 & 8.3 & 8.56E-03 & 15 & 8.2 & 7.16E-03 & 15  & 7.3 & 5.61E-03 & 19  & 9.1 & 7.42E-03 & 7.7 & 7.9 \\
120--150 &  131.1 & && & && & && & && & 1.53E-03 & 17 & 7.9 \\
\end{tabular}
}
\end{sidewaystable}

\clearpage
\begin{sidewaystable}[htb]
\centering
\topcaption{Double-differential cross section times the dimuon branching fraction of the \Pgy\ meson for different ranges of \pt,
in bins of $\abs{y}$ and for the full $\abs{y}$ range, for the unpolarized decay hypothesis, with their statistical and
systematic uncertainties in percent. The average \pt value in each bin is also given. The global uncertainty in
the integrated luminosity of 2.3\% is not included in the systematic uncertainties.
}
\label{tab:psipxsec}
{\small
\begin{tabular}{cc|ccc|ccc|ccc|ccc|ccc}
                            &                         &  \multicolumn{15}{c}{${\cal B} \, \rd\sigma^2/\rd{}\pt \rd{}y$}  \\
\multicolumn{1}{c}{\pt}  & $\langle{\pt}\rangle$ & \multicolumn{3}{c}{$\abs{y} < 0.3$}  & \multicolumn{3}{c}{$0.3 < \abs{y} < 0.6$}  &  \multicolumn{3}{c}{$0.6 < \abs{y} < 0.9$}  & \multicolumn{3}{c}{$0.9 < \abs{y} < 1.2$} & \multicolumn{3}{c}{$\abs{y} < 1.2$} \\
\cline{3-17}
[\GeVns{}]          &  [\GeVns{}]  & [\pbGeV{}]& stat \%  & syst \% & [\pbGeV{}]& stat \%  & syst \% & [\pbGeV{}]& stat \%  & syst \% & [\pbGeV{}]& stat \%  & syst \% & [\pbGeV{}]& stat \%  & syst \% \\
\hline
 20--22 &  21.1 & 1.62E+00 & 3.3 & 5.4 & 1.65E+00 & 3.3 & 5.5 & 1.67E+00 & 3.2 & 5.0 & 1.58E+00 & 3.6 & 5.8 & 1.63E+00 & 1.6 & 5.8 \\
 22--25 &  23.7 & 9.46E-01 & 4.2 & 5.0 & 8.90E-01 & 4.4 & 5.0 & 1.03E+00 & 3.9 & 4.6 & 8.30E-01 & 4.8 & 5.3 & 9.19E-01 & 2.1 & 5.4 \\
 25--28 &  26.2 & 5.23E-01 & 5.6 & 5.0 & 5.12E-01 & 5.7 & 5.2 & 5.23E-01 & 5.3 & 4.4 & 4.89E-01 & 6.1 & 6.7 & 5.10E-01 & 2.8 & 6.8 \\
 28--30 &  28.7 & 3.45E-01 & 6.9 & 5.4 & 3.77E-01 & 6.3 & 5.6 & 3.08E-01 & 7.3 & 5.0 & 3.55E-01 & 6.8 & 6.7 & 3.45E-01 & 3.3 & 6.8 \\
 30--35 &  32.2 & 1.94E-01 & 6.4 & 6.1 & 2.04E-01 & 6.2 & 7.0 & 1.82E-01 & 6.6 & 6.2 & 1.46E-01 & 8.2 & 11  & 1.80E-01 & 3.3 & 11 \\
 35--40 &  37.2 & 9.68E-02 & 9.2 & 7.3 & 8.87E-02 & 9.1 & 8.1 & 8.42E-02 & 9.2 & 9.0 & 7.22E-02 & 11  & 12  & 8.46E-02 & 4.7 & 12 \\
 40--55 &  45.7 & 2.93E-02 & 9.3 & 7.5 & 2.87E-02 & 9.4 & 8.2 & 2.90E-02 & 9.6 & 9.1 & 2.63E-02 & 11 & 12 & 2.81E-02 & 4.8 & 12 \\
 55--75 &  62.5 & 4.75E-03 & 66 & 12 & 5.07E-03 & 21 & 15 & 4.82E-03 & 10 & 15 & 6.59E-03 & 18 & 17 & 4.97E-03 & 11 & 16 \\
 75--100 &  84.2 & 1.01E-03 & 48 & 18 & 1.20E-03 & 17 & 17 & 1.24E-03 & 23 & 18 & 8.81E-04 & 23 & 20 & 1.08E-03 & 6.6 & 20 \\
100--130 &  111.0 & && & && & && & && & 2.85E-04 & 24 & 20 \\
\end{tabular}
}
\end{sidewaystable}

\clearpage
\begin{sidewaystable}[htb]
\centering
\topcaption{Double-differential cross section times the dimuon branching fraction of the \PgUa\ meson for different ranges of \pt,
in bins of $\abs{y}$ and for the full $\abs{y}$ range, for the unpolarized decay hypothesis, with their statistical and
systematic uncertainties in percent. The average \pt value in each bin is also given. The global uncertainty in
the integrated luminosity of 2.3\% is not included in the systematic uncertainties.
}
\label{tab:y1sxsec_eta1}
\begin{tabular}{cc|ccc|ccc|ccc}
\                          &                          & \multicolumn{9}{c}{${\cal B} \, \rd\sigma^2/\rd{}\pt \rd{}y$}  \\
\multicolumn{1}{c}{\pt}  & $\langle{\pt}\rangle$  & \multicolumn{3}{c}{$\abs{y} < 0.6$}  & \multicolumn{3}{c}{$0.6 < \abs{y} < 1.2$}  &  \multicolumn{3}{c}{$\abs{y} < 1.2$} \\
\cline{3-11}
[\GeVns{}]          &  [\GeVns{}]  & [\pbGeV{}]& stat \%  & syst \% & [\pbGeV{}]& stat \%  & syst \% & [\pbGeV{}]& stat \%  & syst \% \\
\hline
 20--22 &  20.9 & 5.76E+00 & 1.7 & 7.1 & 5.46E+00 & 1.7 & 7.8 & 5.62E+00 & 0.9 & 7.3 \\
 22--24 &  22.9 & 3.60E+00 & 1.6 & 6.0 & 3.77E+00 & 2.1 & 6.7 & 3.68E+00 & 1.1 & 6.3 \\
 24--26 &  25.0 & 2.50E+00 & 1.9 & 5.3 & 2.44E+00 & 2.1 & 6.1 & 2.47E+00 & 1.3 & 5.6 \\
 26--28 &  26.9 & 1.72E+00 & 2.2 & 4.8 & 1.75E+00 & 2.5 & 5.3 & 1.73E+00 & 1.5 & 4.9 \\
 28--30 &  29.0 & 1.19E+00 & 2.6 & 4.9 & 1.21E+00 & 3.1 & 5.2 & 1.20E+00 & 1.8 & 4.9 \\
 30--32 &  31.0 & 8.50E-01 & 3.5 & 4.9 & 8.62E-01 & 3.5 & 5.3 & 8.55E-01 & 2.1 & 5.1 \\
 32--34 &  33.0 & 6.43E-01 & 4.0 & 4.9 & 6.26E-01 & 4.1 & 5.0 & 6.36E-01 & 2.4 & 4.8 \\
 34--36 &  35.0 & 4.88E-01 & 4.2 & 4.7 & 4.55E-01 & 5.1 & 5.4 & 4.72E-01 & 2.8 & 5.3 \\
 36--38 &  37.0 & 3.52E-01 & 5.2 & 4.8 & 3.12E-01 & 6.0 & 6.1 & 3.32E-01 & 3.4 & 5.8 \\
 38--40 &  39.0 & 2.37E-01 & 6.1 & 5.0 & 2.65E-01 & 6.5 & 6.2 & 2.51E-01 & 3.9 & 6.0 \\
 40--43 &  41.4 & 2.17E-01 & 5.4 & 5.4 & 1.91E-01 & 5.7 & 5.1 & 2.04E-01 & 3.5 & 4.8 \\
 43--46 &  44.4 & 1.39E-01 & 6.0 & 5.5 & 1.17E-01 & 7.3 & 5.0 & 1.28E-01 & 4.3 & 4.9 \\
 46--50 &  47.9 & 7.87E-02 & 7.0 & 5.2 & 8.25E-02 & 7.0 & 5.1 & 8.07E-02 & 4.6 & 4.6 \\
 50--55 &  52.3 & 5.36E-02 & 7.4 & 5.0 & 4.96E-02 & 8.2 & 5.2 & 5.16E-02 & 5.2 & 4.7 \\
 55--60 &  57.3 & 3.72E-02 & 8.8 & 5.4 & 2.96E-02 & 11 & 5.6 & 3.33E-02 & 6.5 & 4.8 \\
 60--70 &  64.3 & 1.51E-02 & 10 & 5.2 & 1.72E-02 & 11 & 5.6 & 1.62E-02 & 6.8 & 5.1 \\
 70--100 &  80.5 & 3.83E-03 & 12 & 5.4 & 4.85E-03 & 12 & 5.6 & 4.32E-03 & 7.7 & 4.7 \\
100--130 &  111.5 & && & && & 6.48E-04 & 20 & 4.7 \\
\end{tabular}
\end{sidewaystable}

\clearpage
\begin{sidewaystable}[htb]
\centering
\topcaption{Double-differential cross section times the dimuon branching fraction of the \PgUb\ meson for different ranges of \pt,
in bins of $\abs{y}$ and for the full $\abs{y}$ range, for the unpolarized decay hypothesis, with their statistical and
systematic uncertainties in percent. The average \pt value in each bin is also given. The global uncertainty in
the integrated luminosity of 2.3\% is not included in the systematic uncertainties.
}
\label{tab:y2sxsec_eta1}
\begin{tabular}{cc|ccc|ccc|ccc}
\                          &                          & \multicolumn{9}{c}{${\cal B} \, \rd\sigma^2/\rd{}\pt \rd{}y$}  \\
\multicolumn{1}{c}{\pt}  & $\langle{\pt}\rangle$  & \multicolumn{3}{c}{$\abs{y} < 0.6$}  & \multicolumn{3}{c}{$0.6 < \abs{y} < 1.2$}  &  \multicolumn{3}{c}{$\abs{y} < 1.2$} \\
\cline{3-11}
[\GeVns{}]          &  [\GeVns{}]  & [\pbGeV{}]& stat \%  & syst \% & [\pbGeV{}]& stat \%  & syst \% & [\pbGeV{}]& stat \%  & syst \% \\
\hline
 20--22 &  20.9 & 2.45E+00 & 2.6 & 6.1 & 2.30E+00 & 2.7 & 6.9 & 2.38E+00 & 1.4 & 6.4 \\
 22--24 &  22.9 & 1.55E+00 & 2.3 & 5.7 & 1.60E+00 & 3.3 & 7.1 & 1.57E+00 & 1.7 & 6.8 \\
 24--26 &  25.0 & 1.15E+00 & 2.7 & 5.2 & 1.16E+00 & 3.1 & 6.1 & 1.16E+00 & 1.9 & 5.8 \\
 26--28 &  26.9 & 7.54E-01 & 3.3 & 4.9 & 8.27E-01 & 3.7 & 5.9 & 7.89E-01 & 2.3 & 5.6 \\
 28--30 &  29.0 & 5.51E-01 & 3.7 & 5.0 & 5.39E-01 & 4.6 & 5.8 & 5.45E-01 & 2.7 & 5.5 \\
 30--32 &  31.0 & 4.02E-01 & 4.8 & 5.4 & 4.28E-01 & 5.2 & 6.5 & 4.14E-01 & 3.1 & 6.3 \\
 32--34 &  33.0 & 3.04E-01 & 5.8 & 6.0 & 3.31E-01 & 5.8 & 5.7 & 3.18E-01 & 3.6 & 5.5 \\
 34--36 &  35.0 & 2.36E-01 & 5.9 & 4.8 & 2.47E-01 & 7.6 & 6.2 & 2.41E-01 & 4.1 & 6.0 \\
 36--38 &  37.0 & 1.54E-01 & 7.2 & 4.7 & 1.61E-01 & 8.3 & 6.0 & 1.57E-01 & 5.0 & 5.6 \\
 38--40 &  39.0 & 1.28E-01 & 7.8 & 4.9 & 1.23E-01 & 9.4 & 5.8 & 1.26E-01 & 5.7 & 5.6 \\
 40--43 &  41.4 & 9.52E-02 & 7.6 & 5.2 & 1.03E-01 & 8.0 & 5.4 & 9.89E-02 & 5.0 & 5.1 \\
 43--46 &  44.4 & 6.83E-02 & 8.3 & 4.9 & 5.68E-02 & 11 & 5.0 & 6.19E-02 & 6.4 & 4.7 \\
 46--50 &  47.9 & 4.53E-02 & 9.1 & 4.8 & 3.32E-02 & 12 & 5.4 & 3.93E-02 & 6.9 & 5.0 \\
 50--55 &  52.3 & 2.81E-02 & 9.9 & 5.1 & 2.57E-02 & 12 & 5.4 & 2.68E-02 & 7.3 & 4.9 \\
 55--60 &  57.3 & 2.00E-02 & 11 & 5.5 & 1.77E-02 & 14 & 5.8 & 1.88E-02 & 8.6 & 5.1 \\
 60--70 &  64.3 & 8.99E-03 & 13 & 5.5 & 1.10E-02 & 13 & 6.3 & 9.96E-03 & 8.8 & 5.5 \\
 70--100 &  80.5 & 1.91E-03 & 19 & 6.3 & 2.24E-03 & 17 & 7.1 & 2.08E-03 & 12 & 5.9 \\
100--130 &  111.5 & && & && & 5.79E-04 & 35 & 6.0 \\
\end{tabular}
\end{sidewaystable}

\clearpage
\begin{sidewaystable}[htb]
\centering
\topcaption{Double-differential cross section times the dimuon branching fraction of the \PgUc\ meson for different ranges of \pt,
 in bins of $\abs{y}$ and for the full $\abs{y}$ range, for the unpolarized decay hypothesis, with their statistical and
 systematic uncertainties in percent. The average \pt value in each bin is also given.
The global uncertainty in the integrated luminosity of 2.3\% is not included  in the systematic uncertainties.
}
\label{tab:y3sxsec_eta1}
\begin{tabular}{cc|ccc|ccc|ccc}
\                          &                          & \multicolumn{9}{c}{${\cal B} \, \rd\sigma^2/\rd{}\pt \rd{}y$}  \\
\multicolumn{1}{c}{\pt}  & $\langle{\pt}\rangle$  & \multicolumn{3}{c}{$\abs{y} < 0.6$}  & \multicolumn{3}{c}{$0.6 < \abs{y} < 1.2$}  &  \multicolumn{3}{c}{$\abs{y} < 1.2$} \\
\cline{3-11}
[\GeVns{}]          &  [\GeVns{}]  & [\pbGeV{}]& stat \%  & syst \% & [\pbGeV{}]& stat \%  & syst \% & [\pbGeV{}]& stat \%  & syst \% \\
\hline
 20--22 &  20.9 & 1.64E+00 & 3.3 & 7.3 & 1.55E+00 & 3.5 & 7.1 & 1.60E+00 & 1.8 & 6.7 \\
 22--24 &  22.9 & 1.08E+00 & 2.9 & 6.0 & 1.21E+00 & 4.0 & 6.5 & 1.13E+00 & 2.0 & 6.2 \\
 24--26 &  25.0 & 7.94E-01 & 3.4 & 6.2 & 7.81E-01 & 4.1 & 6.4 & 7.88E-01 & 2.4 & 6.1 \\
 26--28 &  26.9 & 5.72E-01 & 3.9 & 6.2 & 5.84E-01 & 4.8 & 6.6 & 5.79E-01 & 2.8 & 6.3 \\
 28--30 &  29.0 & 4.11E-01 & 4.5 & 5.8 & 3.79E-01 & 6.0 & 7.0 & 3.96E-01 & 3.3 & 6.8 \\
 30--32 &  31.0 & 2.93E-01 & 6.0 & 6.4 & 3.10E-01 & 6.6 & 7.3 & 3.01E-01 & 3.8 & 7.0 \\
 32--34 &  33.0 & 2.54E-01 & 6.9 & 6.1 & 2.13E-01 & 7.9 & 6.9 & 2.34E-01 & 4.3 & 6.8 \\
 34--36 &  35.0 & 1.82E-01 & 7.0 & 5.9 & 1.84E-01 & 9.7 & 8.1 & 1.82E-01 & 4.7 & 7.9 \\
 36--38 &  37.0 & 1.41E-01 & 7.9 & 5.8 & 1.27E-01 & 10 & 8.2 & 1.34E-01 & 5.4 & 8.0 \\
 38--40 &  39.0 & 1.16E-01 & 8.7 & 6.0 & 1.23E-01 & 11 & 8.0 & 1.19E-01 & 5.7 & 7.9 \\
 40--43 &  41.4 & 7.36E-02 & 9.4 & 6.4 & 7.23E-02 & 10 & 6.5 & 7.33E-02 & 6.1 & 6.1 \\
 43--46 &  44.4 & 4.46E-02 & 11 & 6.3 & 4.81E-02 & 13 & 6.0 & 4.59E-02 & 7.7 & 5.8 \\
 46--50 &  47.9 & 3.64E-02 & 11 & 6.8 & 2.84E-02 & 13 & 6.4 & 3.24E-02 & 7.7 & 6.0 \\
 50--55 &  52.3 & 2.22E-02 & 12 & 6.9 & 2.28E-02 & 13 & 7.2 & 2.24E-02 & 8.4 & 6.9 \\
 55--60 &  57.3 & 1.48E-02 & 14 & 8.1 & 1.35E-02 & 18 & 8.8 & 1.43E-02 & 11 & 8.3 \\
 60--70 &  64.3 & 6.34E-03 & 16 & 10 & 8.50E-03 & 17 & 11 & 7.36E-03 & 10 & 10 \\
 70--100 &  80.5 & 1.88E-03 & 17 & 13 & 2.10E-03 & 19 & 13 & 1.98E-03 & 12 & 13 \\
100--130 &  111.5 & && & && & 2.30E-04 & 48 & 13 \\
\end{tabular}
\end{sidewaystable}

\clearpage

\begin{sidewaystable}[htb]
\centering
\topcaption{Multiplicative scaling factors to obtain the \JPsi\ differential cross sections for
different polarization scenarios ($\ltth = +1, k, -1$) from the unpolarized cross section
measurements given in Table~\ref{tab:jpsixsec}. The value of $k$ is taken equal to $+0.10$, and corresponds to an
average over \pt of the CMS measurement \cite{cmsjpol}.}
\label{tab:poljpsi}
\begin{tabular}{c|ccc|ccc|ccc|ccc|ccc}
\multicolumn{1}{c|}{\pt} & \multicolumn{3}{c}{$\abs{y}<0.3$} & \multicolumn{3}{c}{$0.3 < \abs{y}< 0.6$} & \multicolumn{3}{c}{$0.6 < \abs{y}< 0.9$} & \multicolumn{3}{c}{$0.9 < \abs{y}< 1.2$} & \multicolumn{3}{c}{$\abs{y}< 1.2$}\\
\cline{2-16}
 [\GeVns{}] & \ltt=+1  & \ltt=$k$ & \ltt=$-$1 & \ltt=+1  & \ltt=$k$ & \ltt=$-$1 & \ltt=+1  & \ltt=$k$ & \ltt=$-$1 & \ltt=+1  & \ltt=$k$ & \ltt=$-$1 & \ltt=+1  & \ltt=$k$ & \ltt=$-$1\\
\hline
20--21 & 1.20 & 1.02  & 0.75 & 1.17  & 1.01 & 0.75 & 1.18  & 1.02 & 0.76 & 1.18  & 1.02 & 0.76 & 1.19  & 1.02 & 0.76 \\
21--22 & 1.19 & 1.02  & 0.76 & 1.17  & 1.01 & 0.76 & 1.18  & 1.02 & 0.77 & 1.17  & 1.02 & 0.77 & 1.18  & 1.02 & 0.76 \\
22--23 & 1.19 & 1.02  & 0.76 & 1.16  & 1.01 & 0.77 & 1.17  & 1.02 & 0.78 & 1.17  & 1.02 & 0.78 & 1.17  & 1.02 & 0.77 \\
23--24 & 1.18 & 1.02  & 0.77 & 1.16  & 1.01 & 0.77 & 1.16  & 1.02 & 0.78 & 1.16  & 1.02 & 0.78 & 1.16  & 1.02 & 0.78 \\
24--25 & 1.17 & 1.02  & 0.77 & 1.15  & 1.01 & 0.78 & 1.16  & 1.02 & 0.79 & 1.15  & 1.02 & 0.79 & 1.16  & 1.02 & 0.78 \\
25--26 & 1.17 & 1.02  & 0.78 & 1.15  & 1.01 & 0.78 & 1.15  & 1.02 & 0.79 & 1.15  & 1.02 & 0.79 & 1.15  & 1.02 & 0.79 \\
26--27 & 1.16 & 1.02  & 0.78 & 1.14  & 1.01 & 0.79 & 1.15  & 1.02 & 0.79 & 1.14  & 1.02 & 0.80 & 1.15  & 1.02 & 0.79 \\
27--28 & 1.16 & 1.02  & 0.79 & 1.14  & 1.01 & 0.79 & 1.14  & 1.02 & 0.80 & 1.14  & 1.02 & 0.81 & 1.14  & 1.02 & 0.80 \\
28--29 & 1.15 & 1.02  & 0.79 & 1.13  & 1.01 & 0.80 & 1.13  & 1.02 & 0.81 & 1.14  & 1.02 & 0.81 & 1.14  & 1.01 & 0.80 \\
29--30 & 1.14 & 1.02  & 0.80 & 1.13  & 1.01 & 0.80 & 1.13  & 1.02 & 0.81 & 1.13  & 1.02 & 0.81 & 1.13  & 1.01 & 0.81 \\
30--32 & 1.14 & 1.02  & 0.81 & 1.12  & 1.01 & 0.81 & 1.12  & 1.01 & 0.82 & 1.13  & 1.01 & 0.82 & 1.13  & 1.01 & 0.81 \\
32--34 & 1.13 & 1.02  & 0.82 & 1.11  & 1.01 & 0.82 & 1.12  & 1.01 & 0.82 & 1.12  & 1.01 & 0.82 & 1.12  & 1.01 & 0.82 \\
34--36 & 1.13 & 1.02  & 0.82 & 1.11  & 1.01 & 0.82 & 1.11  & 1.01 & 0.83 & 1.11  & 1.01 & 0.83 & 1.12  & 1.01 & 0.83 \\
36--38 & 1.11 & 1.01  & 0.83 & 1.11  & 1.01 & 0.83 & 1.11  & 1.01 & 0.84 & 1.10  & 1.01 & 0.85 & 1.11  & 1.01 & 0.84 \\
38--42 & 1.11 & 1.01  & 0.84 & 1.10  & 1.01 & 0.84 & 1.10  & 1.01 & 0.85 & 1.10  & 1.01 & 0.85 & 1.10  & 1.01 & 0.84 \\
42--46 & 1.10 & 1.01  & 0.85 & 1.09  & 1.01 & 0.85 & 1.09  & 1.01 & 0.86 & 1.09  & 1.01 & 0.85 & 1.09  & 1.01 & 0.85 \\
46--50 & 1.09 & 1.01  & 0.86 & 1.08  & 1.01 & 0.86 & 1.08  & 1.01 & 0.87 & 1.09  & 1.01 & 0.86 & 1.09  & 1.01 & 0.86 \\
50--60 & 1.08 & 1.01  & 0.88 & 1.06  & 1.01 & 0.89 & 1.07  & 1.01 & 0.88 & 1.08  & 1.01 & 0.88 & 1.07  & 1.01 & 0.88 \\
60--75 & 1.06 & 1.01  & 0.89 & 1.05  & 1.00 & 0.90 & 1.06  & 1.01 & 0.90 & 1.06  & 1.01 & 0.90 & 1.06  & 1.01 & 0.90 \\
75--95 & 1.05 & 1.01  & 0.91 & 1.04  & 1.00 & 0.93 & 1.04  & 1.01 & 0.92 & 1.05  & 1.01 & 0.91 & 1.04  & 1.01 & 0.92 \\
95--120 & 1.03 & 1.01  & 0.93 & 1.02  & 1.00 & 0.94 & 1.03  & 1.00 & 0.94 & 1.04  & 1.01 & 0.92 & 1.03  & 1.00 & 0.93 \\
120--150 & & & & & & & & & & & & & 1.03  & 1.00 & 0.93 \\
\end{tabular}
\end{sidewaystable}

\begin{sidewaystable}[htb]
\centering
\topcaption{Multiplicative scaling factors to obtain the \Pgy\ differential cross sections for
different polarization scenarios ($\ltth = +1, k, -1$) from the unpolarized cross section
measurements given in Table~\ref{tab:jpsixsec}. The value of $k$ is taken equal to  $+0.03$, and corresponds to an
average over \pt of the CMS measurement \cite{cmsjpol}.}
\label{tab:polpsip}
\begin{tabular}{c|ccc|ccc|ccc|ccc|ccc}
\multicolumn{1}{c|}{\pt} & \multicolumn{3}{c}{$\abs{y}<0.3$} & \multicolumn{3}{c}{$0.3 < \abs{y}< 0.6$} & \multicolumn{3}{c}{$0.6 < \abs{y}< 0.9$} & \multicolumn{3}{c}{$0.9 < \abs{y}< 1.2$} & \multicolumn{3}{c}{$\abs{y}< 1.2$}\\
\cline{2-16}
 [\GeVns{}] & \ltt=+1  & \ltt=$k$ & \ltt=$-$1 & \ltt=+1  & \ltt=$k$ & \ltt=$-$1 & \ltt=+1  & \ltt=$k$ & \ltt=$-$1 & \ltt=+1  & \ltt=$k$ & \ltt=$-$1 & \ltt=+1  & \ltt=$k$ & \ltt=$-$1\\
\hline
20--22 & 1.19 & 1.01  & 0.76 & 1.17  & 0.99 & 0.75 & 1.18  & 1.01 & 0.77 & 1.17  & 1.01 & 0.77 & 1.18  & 1.00 & 0.76 \\
22--25 & 1.18 & 1.01  & 0.77 & 1.15  & 1.00 & 0.77 & 1.16  & 1.01 & 0.78 & 1.16  & 1.01 & 0.78 & 1.16  & 1.00 & 0.78 \\
25--28 & 1.16 & 1.01  & 0.79 & 1.14  & 1.00 & 0.79 & 1.14  & 1.00 & 0.80 & 1.15  & 1.01 & 0.79 & 1.15  & 1.00 & 0.79 \\
28--30 & 1.15 & 1.01  & 0.80 & 1.12  & 1.00 & 0.80 & 1.13  & 1.00 & 0.81 & 1.13  & 1.00 & 0.81 & 1.13  & 1.00 & 0.81 \\
30--35 & 1.14 & 1.01  & 0.81 & 1.12  & 1.00 & 0.81 & 1.12  & 1.00 & 0.82 & 1.12  & 1.00 & 0.82 & 1.12  & 1.00 & 0.82 \\
35--40 & 1.12 & 1.01  & 0.83 & 1.10  & 1.00 & 0.83 & 1.11  & 1.00 & 0.84 & 1.11  & 1.00 & 0.84 & 1.11  & 1.00 & 0.83 \\
40--55 & 1.10 & 1.00  & 0.85 & 1.08  & 1.00 & 0.86 & 1.08  & 1.00 & 0.86 & 1.08  & 1.00 & 0.87 & 1.09  & 1.00 & 0.86 \\
55--75 & 1.07 & 1.00  & 0.88 & 1.07  & 1.00 & 0.88 & 1.06  & 1.00 & 0.89 & 1.06  & 1.00 & 0.89 & 1.07  & 1.00 & 0.88 \\
75--100 & 1.04 & 1.00  & 0.92 & 1.03  & 1.00 & 0.93 & 1.03  & 1.00 & 0.94 & 1.04  & 1.00 & 0.92 & 1.03  & 1.00 & 0.93 \\
100--130 & & & & & & & & & & & & & 1.03  & 1.00 & 0.93 \\
\end{tabular}
\end{sidewaystable}

\begin{table*}[htb]
\centering
\topcaption{Multiplicative scaling factors to obtain the \PgUa\ differential cross sections for
different polarization scenarios ($\ltth = +1, k, -1$) from the unpolarized cross section
measurements given in Table~\ref{tab:y1sxsec_eta1}. The parameter $k$ corresponds to a linear interpolation of the CMS
measured value of $\ltth$~\cite{cmsypol} as a function of \pt for $\pt < 50$\GeV.
For $\pt > 50$\GeV, where no measurements of \ltth\ exist, $k$ is taken
as the average of all the measured values of \ltth\ for $\pt < 50$\GeV.}
\label{tab:polups1}
\begin{tabular}{c|ccc|ccc|ccc}
\multicolumn{1}{c|}{\pt} & \multicolumn{3}{c}{$\abs{y}<0.6$} & \multicolumn{3}{c}{$0.6 < \abs{y}< 1.2$}  & \multicolumn{3}{c}{$\abs{y}< 1.2$}\\
\cline{2-10}
   [\GeVns{}]     & $\ltt=+1$  & $\ltt=k$ & $\ltt=-1$ & $\ltt=+1$  & $\ltt=k$ & $\ltt=-1$ & $\ltt=+1$  & $\ltt=k$ & $\ltt=-1$\\
\hline
20--22 & 1.14 & 0.98  & 0.78 & 1.14  & 0.98 & 0.78 & 1.14  & 0.98 & 0.78  \\
22--24 & 1.13 & 0.99  & 0.78 & 1.13  & 0.99 & 0.78 & 1.13  & 0.99 & 0.78  \\
24--26 & 1.12 & 0.99  & 0.79 & 1.12  & 0.99 & 0.79 & 1.12  & 0.99 & 0.79  \\
26--28 & 1.11 & 0.99  & 0.80 & 1.11  & 0.99 & 0.80 & 1.11  & 0.99 & 0.80  \\
28--30 & 1.11 & 0.99  & 0.81 & 1.11  & 0.99 & 0.81 & 1.11  & 0.99 & 0.81  \\
30--32 & 1.10 & 1.01  & 0.81 & 1.10  & 1.01 & 0.81 & 1.10  & 1.01 & 0.81  \\
32--34 & 1.10 & 1.01  & 0.82 & 1.10  & 1.01 & 0.82 & 1.10  & 1.01 & 0.82  \\
34--36 & 1.09 & 1.01  & 0.82 & 1.09  & 1.01 & 0.82 & 1.09  & 1.01 & 0.82  \\
36--38 & 1.09 & 1.01  & 0.83 & 1.09  & 1.01 & 0.83 & 1.09  & 1.01 & 0.83  \\
38--40 & 1.10 & 1.01  & 0.83 & 1.10  & 1.01 & 0.83 & 1.10  & 1.01 & 0.83  \\
40--43 & 1.08 & 1.01  & 0.84 & 1.08  & 1.01 & 0.84 & 1.08  & 1.01 & 0.84  \\
43--46 & 1.07 & 1.01  & 0.85 & 1.07  & 1.01 & 0.85 & 1.07  & 1.01 & 0.85  \\
46--50 & 1.07 & 1.01  & 0.85 & 1.07  & 1.01 & 0.85 & 1.07  & 1.01 & 0.85  \\
50--55 & 1.06 & 0.99  & 0.86 & 1.06  & 0.99 & 0.86 & 1.06  & 0.99 & 0.86  \\
55--60 & 1.05 & 0.99  & 0.88 & 1.05  & 0.99 & 0.88 & 1.05  & 0.99 & 0.88  \\
60--70 & 1.05 & 0.99  & 0.88 & 1.05  & 0.99 & 0.88 & 1.05  & 0.99 & 0.88  \\
70--100 & 1.03 & 1.00  & 0.92 & 1.03  & 1.00 & 0.92 & 1.03  & 1.00 & 0.92  \\
100--130 & & & & & & & 1.03  & 1.00 & 0.92  \\
\end{tabular}
\end{table*}

\begin{table*}[htb]
\centering
\topcaption{Multiplicative scaling factors to obtain the \PgUb\ differential cross sections for
different polarization scenarios ($\ltth = +1, k, -1$) from the unpolarized cross section
measurements given in Table~\ref{tab:y2sxsec_eta1}. The parameter $k$ corresponds to a linear interpolation of the CMS
measured value of $\ltth$~\cite{cmsypol} as a function of \pt for $\pt < 50$\GeV.
For $\pt > 50$\GeV, where no measurements of \ltth\ exist, $k$ is taken
as the average of all the measured values of \ltth\ for $\pt < 50$\GeV.}
\label{tab:polups2}
\begin{tabular}{c|ccc|ccc|ccc}
\multicolumn{1}{c|}{\pt} & \multicolumn{3}{c}{$\abs{y}<0.6$} & \multicolumn{3}{c}{$0.6 < \abs{y}< 1.2$}  & \multicolumn{3}{c}{$\abs{y}< 1.2$}\\
\cline{2-10}
   [\GeVns{}]     & $\ltt=+1$  & $\ltt=k$ & $\ltt=-1$ & $\ltt=+1$  & $\ltt=k$ & $\ltt=-1$ & $\ltt=+1$  & $\ltt=k$ & $\ltt=-1$\\
\hline
20--22 & 1.14 & 1.03  & 0.78 & 1.14  & 1.03 & 0.78 & 1.14  & 1.03 & 0.78  \\
22--24 & 1.13 & 1.03  & 0.79 & 1.13  & 1.03 & 0.79 & 1.13  & 1.03 & 0.79  \\
24--26 & 1.12 & 1.03  & 0.79 & 1.12  & 1.03 & 0.79 & 1.12  & 1.03 & 0.79  \\
26--28 & 1.11 & 1.03  & 0.80 & 1.11  & 1.03 & 0.80 & 1.11  & 1.03 & 0.80  \\
28--30 & 1.11 & 1.03  & 0.81 & 1.11  & 1.03 & 0.81 & 1.11  & 1.03 & 0.81  \\
30--32 & 1.10 & 1.03  & 0.82 & 1.10  & 1.03 & 0.82 & 1.10  & 1.03 & 0.82  \\
32--34 & 1.10 & 1.03  & 0.82 & 1.10  & 1.03 & 0.82 & 1.10  & 1.03 & 0.82  \\
34--36 & 1.09 & 1.03  & 0.82 & 1.09  & 1.03 & 0.82 & 1.09  & 1.03 & 0.82  \\
36--38 & 1.09 & 1.03  & 0.83 & 1.09  & 1.03 & 0.83 & 1.09  & 1.03 & 0.83  \\
38--40 & 1.09 & 1.03  & 0.83 & 1.09  & 1.03 & 0.83 & 1.09  & 1.03 & 0.83  \\
40--43 & 1.08 & 1.03  & 0.84 & 1.08  & 1.03 & 0.84 & 1.08  & 1.03 & 0.84  \\
43--46 & 1.07 & 1.02  & 0.85 & 1.07  & 1.02 & 0.85 & 1.07  & 1.02 & 0.85  \\
46--50 & 1.07 & 1.02  & 0.86 & 1.07  & 1.02 & 0.86 & 1.07  & 1.02 & 0.86  \\
50--55 & 1.06 & 0.99  & 0.87 & 1.06  & 0.99 & 0.87 & 1.06  & 0.99 & 0.87  \\
55--60 & 1.06 & 0.99  & 0.86 & 1.06  & 0.99 & 0.86 & 1.06  & 0.99 & 0.86  \\
60--70 & 1.05 & 0.99  & 0.90 & 1.05  & 0.99 & 0.90 & 1.05  & 0.99 & 0.90  \\
70--100 & 1.03 & 0.99  & 0.92 & 1.03  & 0.99 & 0.92 & 1.03  & 0.99 & 0.92  \\
100--130 & & & & & & & 1.03  & 0.99 & 0.92  \\
\end{tabular}
\end{table*}

\begin{table*}[htb]
\centering
\topcaption{Multiplicative scaling factors to obtain the \PgUc\ differential cross sections for
different polarization scenarios ($\ltth = +1, k, -1$) from the unpolarized cross section
measurements given in Table~\ref{tab:y3sxsec_eta1}. The parameter $k$ corresponds to a linear interpolation of the CMS
measured value of $\ltth$~\cite{cmsypol} as a function of \pt for $\pt < 50$\GeV.
For $\pt > 50$\GeV, where no measurements of \ltth\ exist, $k$ is taken
as the average of all the measured values of \ltth\ for $\pt < 50$\GeV, which are all
consistent with a single value.}
\label{tab:polups3}
\begin{tabular}{c|ccc|ccc|ccc}
\multicolumn{1}{c|}{\pt} & \multicolumn{3}{c}{$\abs{y}<0.6$} & \multicolumn{3}{c}{$0.6 < \abs{y}< 1.2$}  & \multicolumn{3}{c}{$\abs{y}< 1.2$}\\
\cline{2-10}
   [\GeVns{}]     & $\ltt=+1$  & $\ltt=k$ & $\ltt=-1$ & $\ltt=+1$  & $\ltt=k$ & $\ltt=-1$ & $\ltt=+1$  & $\ltt=k$ & $\ltt=-1$\\
\hline
20--22 & 1.13 & 1.03  & 0.78 & 1.13  & 1.03 & 0.78 & 1.13  & 1.03 & 0.78  \\
22--24 & 1.13 & 1.02  & 0.79 & 1.13  & 1.02 & 0.79 & 1.13  & 1.02 & 0.79  \\
24--26 & 1.12 & 1.02  & 0.79 & 1.12  & 1.02 & 0.79 & 1.12  & 1.02 & 0.79  \\
26--28 & 1.11 & 1.02  & 0.80 & 1.11  & 1.02 & 0.80 & 1.11  & 1.02 & 0.80  \\
28--30 & 1.11 & 1.02  & 0.81 & 1.11  & 1.02 & 0.81 & 1.11  & 1.02 & 0.81  \\
30--32 & 1.10 & 1.03  & 0.82 & 1.10  & 1.03 & 0.82 & 1.10  & 1.03 & 0.82  \\
32--34 & 1.10 & 1.03  & 0.82 & 1.10  & 1.03 & 0.82 & 1.10  & 1.03 & 0.82  \\
34--36 & 1.09 & 1.03  & 0.83 & 1.09  & 1.03 & 0.83 & 1.09  & 1.03 & 0.83  \\
36--38 & 1.09 & 1.03  & 0.83 & 1.09  & 1.03 & 0.83 & 1.09  & 1.03 & 0.83  \\
38--40 & 1.09 & 1.03  & 0.84 & 1.09  & 1.03 & 0.84 & 1.09  & 1.03 & 0.84  \\
40--43 & 1.08 & 1.03  & 0.84 & 1.08  & 1.03 & 0.84 & 1.08  & 1.03 & 0.84  \\
43--46 & 1.07 & 1.02  & 0.85 & 1.07  & 1.02 & 0.85 & 1.07  & 1.02 & 0.85  \\
46--50 & 1.06 & 1.02  & 0.86 & 1.06  & 1.02 & 0.86 & 1.06  & 1.02 & 0.86  \\
50--55 & 1.06 & 0.99  & 0.87 & 1.06  & 0.99 & 0.87 & 1.06  & 0.99 & 0.87  \\
55--60 & 1.06 & 0.99  & 0.87 & 1.06  & 0.99 & 0.87 & 1.06  & 0.99 & 0.87  \\
60--70 & 1.05 & 0.99  & 0.89 & 1.05  & 0.99 & 0.89 & 1.05  & 0.99 & 0.89  \\
70--100 & 1.03 & 0.99  & 0.92 & 1.03  & 0.99 & 0.92 & 1.03  & 0.99 & 0.92  \\
100--130 & & & & & & & 1.03  & 0.99 & 0.92  \\
\end{tabular}
\end{table*}

\clearpage

\begin{sidewaystable}[htb]
\centering
\topcaption{
Ratios of the \pt differential cross sections times dimuon branching fractions of the prompt
\Pgy\ to \JPsi , \PgUb\ to \PgUa, and \PgUc\ to \PgUa\ mesons for $\abs{y} < 1.2$, with
their statistical and systematic uncertainties in percent.
}
\label{tab:ratio_excited}
\begin{tabular}{c|ccc|c|c|ccc|ccc}
\multicolumn{1}{c|}{\pt} & \multicolumn{3}{c|}{\Pgy\ / \JPsi} & & \multicolumn{1}{c|}{\pt} & \multicolumn{3}{c|}{\PgUb / \PgUa} & \multicolumn{3}{c}{\PgUc / \PgUa}\\
\cline{2-4}\cline{7-12}
\multicolumn{1}{c|}{[\GeVns{}]} & & stat \% & syst \% & &  \multicolumn{1}{c|}{[\GeVns{}]} & & stat \% & syst \% & & stat \% & syst \%\\
\cline{1-4}\cline{6-12}
20--21   & 0.04 & 3 &  11 &   & 20--22   & 0.42 & 1.7 & 8.5  & 0.28 & 2.0 & 9.1 \\
21--22   & 0.04 & 3 &  11 &   & 22--24   & 0.43 & 2.0 & 8.4  & 0.31 & 2.3 & 7.5 \\
22--23   & 0.04 & 3 &  11 &   & 24--26   & 0.47 & 2.3 & 6.9  & 0.32 & 2.7 & 7.1 \\
23--24   & 0.05 & 3 &  11 &   & 26--28   & 0.46 & 2.8 & 6.2  & 0.33 & 3.1 & 6.8 \\
24--25   & 0.04 & 4 &  13 &   & 28--30   & 0.45 & 3.2 & 6.4  & 0.33 & 3.8 & 8.1 \\
25--26   & 0.04 & 4 &  14 &   & 30--32   & 0.48 & 3.8 & 7.8  & 0.35 & 4.3 & 8.9 \\
26--27   & 0.04 & 5 &  15 &   & 32--34   & 0.50 & 4.3 & 6.3  & 0.37 & 4.9 & 8.3 \\
27--28   & 0.04 & 5 &  16 &   & 34--36   & 0.51 & 5.0 & 7.9  & 0.38 & 5.5 & 11 \\
28--29   & 0.05 & 5 &  17 &   & 36--38   & 0.47 & 6.1 & 8.3  & 0.40 & 6.4 & 11 \\
29--30   & 0.05 & 6 &  17 &   & 38--40   & 0.50 & 6.8 & 8.3  & 0.47 & 6.9 & 11 \\
30--32   & 0.04 & 5 &  18 &   & 40--43   & 0.48 & 6.1 & 5.7  & 0.36 & 7.0 & 7.1\\
32--34   & 0.04 & 6 &  18 &   & 43--46   & 0.48 & 7.7 & 5.4  & 0.36 & 8.8 & 6.6\\
34--36   & 0.04 & 6 &  20 &   & 46--50   & 0.49 & 8.3 & 5.3  & 0.40 & 9.0 & 6.6\\
36--38   & 0.04 & 7 &  20 &   & 50--55   & 0.52 & 8.9 & 5.4  & 0.43 & 9.8 & 7.5 \\
38--42   & 0.05 & 6 &  21 &   & 55--60   & 0.56 & 11  & 6.2  & 0.43 & 12  & 9.2 \\
42--46   & 0.04 & 9 &  22 &   & 60--70   & 0.62 & 11  & 6.6  & 0.46 & 12  & 11 \\
46--50   & 0.05 & 10 & 25 &   & 70--100  & 0.48 & 14  & 6.6  & 0.46 & 14  & 14 \\
50--60   & 0.05 & 10 & 25 &   & 100--130 & 0.89 & 40  & 6.6  & 0.35 & 52  & 14 \\
60--75   & 0.04 & 42 & 26 &   & & & & & & & \\
75--95   & 0.05 & 10 & 29 &   & & & & & & & \\
95--120  & 0.06 & 21 & 38 &   & & & & & & & \\
120--150 & 0.07 & 71 & 38 &   & & & & & & & \\
\end{tabular}
\end{sidewaystable}

}
\cleardoublepage \section{The CMS Collaboration \label{app:collab}}\begin{sloppypar}\hyphenpenalty=5000\widowpenalty=500\clubpenalty=5000\textbf{Yerevan Physics Institute,  Yerevan,  Armenia}\\*[0pt]
A.M.~Sirunyan, A.~Tumasyan
\vskip\cmsinstskip
\textbf{Institut f\"{u}r Hochenergiephysik,  Wien,  Austria}\\*[0pt]
W.~Adam, F.~Ambrogi, E.~Asilar, T.~Bergauer, J.~Brandstetter, E.~Brondolin, M.~Dragicevic, J.~Er\"{o}, M.~Flechl, M.~Friedl, R.~Fr\"{u}hwirth\cmsAuthorMark{1}, V.M.~Ghete, J.~Grossmann, J.~Hrubec, M.~Jeitler\cmsAuthorMark{1}, A.~K\"{o}nig, N.~Krammer, I.~Kr\"{a}tschmer, D.~Liko, T.~Madlener, I.~Mikulec, E.~Pree, D.~Rabady, N.~Rad, H.~Rohringer, J.~Schieck\cmsAuthorMark{1}, R.~Sch\"{o}fbeck, M.~Spanring, D.~Spitzbart, W.~Waltenberger, J.~Wittmann, C.-E.~Wulz\cmsAuthorMark{1}, M.~Zarucki
% ...
\vskip\cmsinstskip
% ...
1:~~Also at Vienna University of Technology, Vienna, Austria\\
% ...

\end{sloppypar}
\end{document}